\documentclass[prb,amsmath,amssymb,superscriptaddress,twocolumn]{revtex4}
\usepackage{amsmath}
\usepackage{braket}
\usepackage{graphicx}
\usepackage{multirow}
\usepackage{bbm,color,ulem}
\DeclareMathAlphabet{\mathpzc}{OT1}{pzc}{m}{it}
\DeclareMathOperator{\sinc}{sinc}
\allowdisplaybreaks
\begin{document}
\title{Crosstalk error correction through dynamical decoupling of single-qubit gates in capacitively coupled singlet-triplet semiconductor spin qubits}
\author{Donovan Buterakos}
\author{Robert E.\ Throckmorton}
\author{S.\ Das Sarma}
\affiliation{Condensed Matter Theory Center and Joint Quantum Institute, Department of Physics, University of Maryland, College Park, Maryland 20742-4111 USA}
\date{\today}
\begin{abstract}
In addition to magnetic field and electric charge noise adversely affecting spin qubit operations, performing
single-qubit gates on one of multiple coupled singlet-triplet qubits presents a new challenge---crosstalk, which
is inevitable (and must be minimized) in any multiqubit quantum computing architecture.  We develop a set of
dynamically-corrected pulse sequences that are designed to cancel the effects of both types of noise (i.e., field
and charge) as well as crosstalk to leading order, and provide parameters for these corrected sequences for all $24$
of the single-qubit Clifford gates.  We then provide an estimate of the error as a function of the noise and capacitive
coupling to compare the fidelity of our corrected gates to their uncorrected versions.  Dynamical error correction
protocols presented in this work are important for the next generation of singlet-triplet qubit devices where coupling
among many qubits will become relevant.
\end{abstract}
\maketitle

\section{Introduction} \label{Sec:Introduction}
Correcting for error in operations on qubits is of utmost importance to building a working quantum computer.
The fact that quantum error correction is possible is what started the whole world-wide  effort in trying to
build a practical quantum computer.  Operations on qubits require very high precision and accuracy; in fact,
a fidelity of at least $99\%$ is required to implement error correction using surface codes\cite{FowlerPRA2012}.
Other error correction techniques require an even higher fidelity of $99.99\%$.  Several different platforms
for realizing qubits exist, but our focus in this work will be on electronic spins in semiconductor quantum
dots.  Several different types of semiconductor quantum dot electron spin qubits exist, such as the single-spin
exchange qubit\cite{LossPRA1998,NowackScience2011,PlaNature2012,PlaNature2013,VeldhorstNatNano2014,BraakmanNatNano2013,OtsukaSciRep2016,ItoSciRep2016},
the singlet-triplet two-electron double-dot qubit\cite{LevyPRL2002,PettaScience2005,FolettiNatPhys2009,VanWeperenPRL2011,MauneNature2012,ShulmanScience2012,DialPRL2013,ShulmanNatCommun2014,ReedPRL2016,MartinsPRL2016}, the exchange-only three-electron triple-dot qubit\cite{DiVincenzoNature2000,MedfordNatNano2013,MedfordPRL2013,EngSciAdv2015,ShimPRB2016},
and the ``hybrid'' three-electron double-dot qubit\cite{ShiPRL2012,KimNature2014,KimNPJQI2015}.  The semiconductor
spin qubit platform has the advantages of being compatible with the existing semiconductor electronics industry
as well as the ability to perform gates more quickly (using fast electrical pulses) than other platforms, such as superconducting
and ion trap qubits.  Other perceived advantages of semiconductor spin qubits include the scalability inherent in semiconductors
and the relatively long spin relaxation times in solids.  However, noise-induced error has been a formidable challenge
adversely affecting experimental progress in spin qubits.  Fortunately, considerable progress has been achieved recently,
with single-qubit gate fidelities of $99\%$ and two-qubit gate fidelities of $90\%$ having been reported in singlet-triplet
double-dot qubits\cite{NicholNPJQI2017}.  Several methods for reducing error have been developed for these platforms.
Methods such as isotopic purification in Si and polarization of nuclei in GaAs reduce noise at the materials level
simply by eliminating the sources of field noise.  Other methods instead combat the effects of noise through pulse
control techniques such as dynamical decoupling, Bayesian estimation of parameters, and designing dynamically corrected pulse
sequences that partially cancel the effects of noise\cite{BluhmPRL2010,BluhmNatPhys2011,SergeevichPRA2011,ShulmanNatCommun2014,MuhonenNatNanotechnol2014,MalinowskiNatNanotechnol2017}.
The goal is to sufficiently reduce both one- and two-qubit gate errors in the physical qubits so that surface code
architectures become feasible, eventually leading to quantum error corrections producing logical qubits.

In spite of enormous progress over the last 5-10 years, the experimental situation in semiconductor spin qubits is
still somewhat discouraging compared with superconducting and ion trap qubits since reasonable two-qubit gate operations
have only been demonstrated in singlet-triplet qubits in GaAs with the fidelity improving from\cite{ShulmanScience2012} $~70\%$
to\cite{NicholNPJQI2017} $~90\%$ over the 2012-2017 five-year period.  Although single-qubit fidelity approaching or
even exceeding $99\%$ has been reported for both single spin and singlet-triplet qubits, these experiments are not carried
out in multiqubit platforms and, therefore, one does not know the limiting single-qubit fidelity in circuits where multiple
qubits are being operated.  Given that the most advanced gate operations by far have happened so far only in the singlet-triplet
spin qubits, our focus in this work is entirely on this system.  The specific issue we address is the mitigation of
crosstalk errors where the quantum computing platform consists of many singlet-triplet qubits with single-qubit gate operations
going on simultaneously in many of them, as would eventually be necessary for any meaningful quantum information processing
task.

There has been considerable work on error correction techniques in semiconductor quantum dot systems.  For
the single-electron exchange qubit, NMR-inspired decoupling techniques such as the Carr-Purcell-Meiboom-Gill (CPMG)
technique, a generalization of the Hahn echo technique\cite{WitzelPRL2007,WitzelPRB2007,LeePRL2008}, exist.  Unfortunately,
these techniques cannot be applied to singlet-triplet qubits, which are the focus of this work.  These generalized
spin-echo-type protocols for restoring quantum coherence require one to be able to apply, for example, a $\pi$ rotation
pulse and then later apply a $-\pi$ rotation pulse about the same axis, which cannot be done in existing
experimental singlet-triplet systems.  To see why, we consider the effective Hamiltonian for a singlet-triplet
qubit within the logical subspace, which is
\begin{equation}
H=J(t)Z+hX,
\end{equation}
where $J(t)$ is the exchange coupling, $h$ is the magnetic field difference between the two qubits, and $X$
and $Z$ are the Pauli matrices.  The magnetic field difference is realized using a micromagnet or by polarizing
the nuclei (if possible).  This difference may in principle also be realized by electrically tuning the effective
$g$ factors in the quantum dots\cite{YokoshiPRB2010,NenashevJApplPhys2015,HwangPRB2017}, but this technique has
not been used in any singlet-triplet qubit experiment to date.  It is very difficult to control this field difference
quickly in the actual experiments, and thus it is held constant.  This means that all control is achieved by electrically
tuning the exchange coupling, either by tilting the dot potential or by changing the height of the potential barrier
between them so that the wave function overlap is modified.  Furthermore, at least in the presence of just two electrons,
it is very difficult to make the exchange coupling negative.  We should note, however, that it can be made negative in
a sufficiently strong magnetic field or in the presence of more than two electrons\cite{HuPRA2001,NielsenPRB2013,MartinsArXiv2017},
though the presence of these additional electrons enlarges the total Hilbert space of the system, creating a new
set of challenges such as leakage errors.  All of this means that we cannot apply the time-reversed version of a
given pulse or pulse sequence in a singlet-triplet qubit.  We must therefore seek other methods for error correction
through dynamical decoupling.

There have been several papers proposing dynamically corrected pulse sequences for performing single-qubit
singlet-triplet gates.  Three papers have been written using the \textsc{supcode} technique, one for the case in which only (electrical)
charge noise is present\cite{WangNatComm2012} and two others in which (magnetic) field noise is also included\cite{WangPRA2014,ThrockmortonArXiv}.
This method, which we will be employing a generalization of in this work, consists of inserting an uncorrected
``identity'' operation into the pulse sequence for a given gate that is arranged in such a way as to cancel
the noise-induced error to leading order.  These works consider sequences of square pulses; another work\cite{ZengArXiv}
considers the case with only field noise and proposes smooth pulses that cancel noise-induced error to arbitrary
orders.  In relation to two-qubit operations, the problem of the dynamics of two coupled singlet-triplet qubits
under the influence of noise has been investigated\cite{DasSarmaPRB2016}, and a recent paper discusses the fidelity
for realizing a maximally-entangled state from a tensor-product state\cite{WuArXiv}.  Dynamical error correction
has also been investigated for controlled-NOT (CNOT) gates, first to a limited extent in Ref. \onlinecite{WangPRA2014} and then in
more detail in Ref. \onlinecite{WangNPJQI2015}.  While correcting for noise-induced error is important, it is not
the only issue that needs to be addressed in the context of practical quantum information processing.  At least
one two-qubit entangling operation, along with the ability to perform arbitrary single-qubit gates, is necessary
to achieve universal quantum computation.  Thus, one must consider also the coupling among the singlet-triplet qubits.

Unfortunately, the couplings between qubits required to perform two-qubit operations also adversely affect single-qubit
operations via crosstalk.  However, as discussed later, the strength of the capacitive coupling is proportional to the
value of the exchange couplings $J$ for each qubit, so one could set $J$ for the neighboring qubit to zero while performing
the single-qubit gate to counter this problem.  Unfortunately, this poses two problems. First, there are experimental limitations in
setting $J$ to exactly zero, and thus some small crosstalk would still persist. Secondly, a dynamical pulse sequence is
required even for qubits on which no gates are performed to maintain coherence. Therefore, in order to be compatible with
a large-scale architecture, it is necessary to address the system as a whole, which is in fact the subject of this work.
Our dynamical decoupling sequence for correcting crosstalk errors (along with field and charge noise) is not an essential
ingredient for the currently ongoing experiments where one has at best only two qubits coupled with each other.  But, this
is of course a very primitive state of affairs as far as quantum computing goes.  Eventually, the platform must consist of
many coupled qubits with substantial crosstalk among them.  Our work will become an important ingredient when both single-
and multiqubit gate operations are being carried out in quantum computing platforms with many coupled singlet-triplet qubits.
Our work is also relevant to a system containing only two singlet-triplet qubits where the exchange coupling within the qubit
cannot simply be tuned to zero while turning on the two-qubit capacitive coupling.

We will be considering here two capacitively coupled singlet-triplet qubits.  This form of coupling, which is used extensively
for inter-qubit coupling in singlet-triplet qubits\cite{ShulmanScience2012,NicholNPJQI2017}, is simpler to treat
than a Heisenberg exchange coupling, since capacitive coupling, as we will see, cannot cause leakage out of the logical
subspace.  We also assume that the field and charge noise in our system is quasistatic.  As noted earlier, our
approach to dynamical error correction is a generalization of the \textsc{supcode} approach of Refs. \onlinecite{WangPRA2014}
and \onlinecite{ThrockmortonArXiv} and, like this approach, the overall idea is to insert an uncorrected ``identity''
operation into the pulse sequence for a given gate that is arranged in such a way as to cancel both noise-
and crosstalk-induced error to leading order.  We begin by deriving expressions giving the noise- and crosstalk-induced
error to leading order for a single pulse (i.e., all Hamiltonian parameters held constant).  From this, we
show how to derive the error for a general, multi-pulse, sequence.  Our error correction procedure consists
of first applying our ``na\"ive'' pulse sequence for a given gate to the qubits, and then following that with
an uncorrected ``identity'' with its parameters arranged in such a way as to cancel both noise- and crosstalk-induced
errors.  This ``identity'' consists of a set of ``blocks'' of pulse sequences such that one qubit is subject to
a single pulse and the other is subject to two pulses with a combined duration equal to that of the single
pulse; we adopt such a form for mathematical simplicity.  We then determine the parameters needed to make the
sum of the errors from the ``na\"ive'' sequence and the ``identity'' zero.  We present the parameters that we
extract from this procedure for all $24$ Clifford gates and present an example, a rotation by $\pi$ about the
$\hat{\vec{x}}+\hat{\vec{y}}$ axis.  Overall, the resulting pulse sequences consist of more ``blocks'' (up to
nine) than those used to correct only noise-induced error for a single qubit in isolation.  The presence of
crosstalk also forces us to propose sequences that never set the exchange couplings to their maximum experimentally
possible values.  This is because the capacitive coupling strength is proportional to the exchange couplings, and
thus large exchange couplings will result in more crosstalk.  We then, as an important check, demonstrate that
these proposed error-corrected sequences do, in fact, cancel noise- and crosstalk-induced errors to leading order.

The rest of the paper is organized as follows.  In Sec.\ II, we present our model and derive the formulas for
the noise- and crosstalk-induced error to first order.  Section\ III details our dynamical error correction scheme,
presents the mathematical form of the corrected pulse sequences in terms of a set of numerical parameters, and
demonstrates that our correction scheme does in fact cancel the leading-order noise- and crosstalk-induced error.
We give our conclusions in Sec.\ IV, and provide the specific numerical values of the parameters for the dynamically-corrected
pulse sequences implementing the Clifford gates in Appendix A. We also provide in Appendix B the dynamical decoupling
pulse sequences for a different model of the inter-qubit coupling, where the coupling is independent of the intra-qubit
exchange energy.

\section{Model}
\subsection{Hamiltonian}

A singlet-triplet (ST) qubit consists of a pair of quantum dot spins coupled by the exchange interaction with a
computational basis given by $\ket{0}=\frac{1}{\sqrt{2}}(\ket{\uparrow\downarrow}+\ket{\downarrow\uparrow})$
and $\ket{1}=\frac{1}{\sqrt{2}}(\ket{\uparrow\downarrow}-\ket{\downarrow\uparrow})$.  Single-qubit control
is obtained by applying a magnetic field gradient $h$ across the two quantum dots and by varying the strength of
the exchange interaction $J$, yielding the single-qubit Hamiltonian $H=hX+JZ$, where $X$ and $Z$ are Pauli matrices in the logical basis.
However, due to physical limitations of the system, these parameters incur several constraints. First, the time scale
on which the magnetic field gradient can be varied is much longer than the time needed to vary the exchange
coupling, so, in practice, $h$ is held constant. Second, the strength of the exchange interaction $J$ must be positive
and, in general, bounded between some values $j_\text{min}$ and $j_\text{max}$. Thus, axes of rotation are limited
to the first quadrant of the $xz$ plane.

We consider a system of two such ST qubits with a large enough potential barrier that there is no exchange interaction
between the two separate qubits, but close enough that a capacitive inter-qubit coupling of strength $J_{12}$ is present
between them.  This coupling arises from the difference in charge distributions between the two quantum dots of a ST qubit.
When a ST qubit is in the singlet state $\ket{1}$, the charge distribution between the two quantum dots is asymmetric,
producing a dipole moment. If instead the qubit is in the triplet state $\ket{0}$, the charge distribution is symmetric,
so no dipole moment is present. The Hamiltonian thus includes a dipole-dipole interaction which contributes only when both
qubits are in the singlet state\cite{ShulmanScience2012}. The strength of this coupling term is proportional to both
$J_1$ and $J_2$, and we will set the proportionality constant $\varepsilon$. Thus, for a system of two capacitively coupled
singlet-triplet qubits we have the following Hamiltonian:
\begin{eqnarray}
H&=&h_1X_1+h_2X_2+J_1Z_1+J_2Z_2\cr
&+&\varepsilon J_1J_2(Z_1-1)(Z_2-1). \label{eqn:basehamiltonian}
\end{eqnarray}
While the capacitive coupling term of the above Hamiltonian has often been quoted as an experimentally-established fact,
and not derived from any model, it is possible to put it on a more rigorous footing as follows.  Let us consider the Hubbard
model Hamiltonian for the double-dot system considered in Ref.\ \onlinecite{LiPRB2012} [Eq.\ (A3), minus the $H_J$ terms,
which are only present when there are more than two electrons in the double-dot system\cite{HuPRA2001,NielsenPRB2013}]:
\begin{equation}
H=H_e+H_t+H_U,
\end{equation}
with
\begin{eqnarray}
H_e&=&\sum_{k=1}^{2}\epsilon_k n_{k}, \\
H_t&=&t\sum_{\sigma=\uparrow,\downarrow}c^{\dagger}_{1\sigma}c_{2\sigma}+\text{H.c.}, \\
H_U&=&U\sum_{k=1}^{2}n_{k\uparrow}n_{k\downarrow}+U'n_1n_2.
\end{eqnarray}
The $c_{k\sigma}$ operators annihilate an electron in dot $k$ with spin $\sigma$, $n_{k\sigma}=c^{\dagger}_{k\sigma}c_{k\sigma}$
is the number of electrons of spin $\sigma$ in dot $k$, and $n_k=\sum_{\sigma}n_{k\sigma}$.
Throughout, we will assume that $U,U'\gg t$.  We can write the dipole moment as
$\vec{d}=\tfrac{1}{2}e\vec{a}(n_2-n_1)$, with $e$ the electron charge and $\vec{a}$ the vector pointing
from the first quantum dot to the second.  If we let $\epsilon_1=\epsilon_0-\tfrac{1}{2}\Delta\epsilon$
and $\epsilon_2=\epsilon_0+\tfrac{1}{2}\Delta\epsilon$, then we may write
\begin{equation}
\vec{d}=e\vec{a}\frac{\partial H}{\partial\Delta\epsilon}.
\end{equation}
By the Feynman-Hellmann theorem, we find that the expectation value of this dipole moment
operator is just
\begin{equation}
\braket{\vec{d}}=e\vec{a}\frac{\partial E}{\partial\Delta\epsilon},
\end{equation}
where $E$ is the (expectation value of the) energy of the system.  The derivative of this
energy will be zero in the triplet state, since the Pauli exclusion principle would
forbid both electrons from occupying the same dot, even if $\epsilon_1\neq\epsilon_2$.
Therefore, only the singlet state will have a dipole moment.  However, in this case, the
energy difference between the singlet and triplet states just gives the exchange coupling
$J$, so that the expectation value of the dipole moment in the singlet state is just
\begin{equation}
\braket{\vec{d}}=e\vec{a}\frac{\partial J}{\partial\Delta\epsilon},
\end{equation}
and is zero in the triplet state.

To establish that $\left |\braket{\vec{d}}\right |\propto J$, we now use Eq.\ (A14)
of Ref.\ \onlinecite{LiPRB2012} (again, minus contributions from the $H_J$ terms),
\begin{equation}
J=\frac{4t^2}{U-U'-\left |\Delta\epsilon\right |}.
\end{equation}
For $\left |\Delta\epsilon\right |\ll U-U'$,
\begin{equation}
\frac{\partial J}{\partial\Delta\epsilon}\approx \frac{4t^2}{(U-U')^2}\mbox{sgn}(\Delta\epsilon)=\frac{J}{U-U'}\mbox{sgn}(\Delta\epsilon),
\end{equation}
and therefore
\begin{equation}
\braket{\vec{d}}\approx \frac{eJ}{U-U'}\mbox{sgn}(\Delta\epsilon)\vec{a}.
\end{equation}
There will be corrections to this formula, but these will be small, as they are proportional
to powers of $\frac{\left |\Delta\epsilon\right |}{U-U'}$.  Experimental data on noise in the
system imply that $\frac{\partial J}{\partial\epsilon}\propto J$, so it seems that, at least
in experimental systems, this approximation works very well qualitatively.

Finally, to obtain the interaction term for two double-dot systems, we may approximate it using
the classical dipole-dipole interaction potential,
\begin{equation}
U_{dd}=\frac{3(\vec{d_1}\cdot\vec{r})(\vec{d_2}\cdot\vec{r})-\vec{d_1}\cdot\vec{d_2}r^2}{r^5},
\end{equation}
where $\vec{r}$ is the vector pointing from one dipole to the other.  We can already see from
our previous result simply by substituting in the expectation values of the dipole moments of
the double dots that this will be proportional to $J_1 J_2$ when both double dots are in
the singlet state, with the proportionality factor being determined by the relative position
and orientation of the double dots, and zero otherwise, thus establishing the capacitive coupling
term given in Eq.\ \eqref{eqn:basehamiltonian}.

We should note that arriving at these expressions required a number of simplifying approximations;
a more detailed analysis of this problem, starting from a microscopic model of the full four-dot
system, would be required to give a fully rigorous justification of this form.  While such a calculation
would be of great interest, it is beyond the scope of this work, where we stick to Eq.\ \eqref{eqn:basehamiltonian}
as describing the inter-qubit coupling Hamiltonian as used extensively in the theoretical literature
on singlet-triplet qubits.

\subsection{Expansion of evolution operator}

In practice\cite{ShulmanScience2012}, $\varepsilon J_i$ is found to be on the order of $\tfrac{1}{300}$, so we choose to approach
this problem by performing a power series expansion in $\varepsilon$. This allows us to consider a simplified Ising
interaction term of the form $\varepsilon J_1J_2Z_1Z_2$, since the terms $-\varepsilon J_1J_2Z_i$ can be absorbed into
the $J_iZ_i$ terms without affecting the result to first order in $\varepsilon$ (see Eq.\ \eqref{eqn:hamiltonianwitherr}).
For convenience, we introduce the shorthand $a_i=\sqrt{h_i^2+J_i^2}$, and perform the expansion in the rotated frame
given by:

\begin{align}
\begin{split}
&X_i'=(h_iX_i+J_iZ_i)/a_i, \\
&Y_i'=Y_i, \\
&Z_i'=(-J_iX_i+h_iZ_i)/a_i.
\label{eqn:paulirotation}
\end{split}
\end{align}

This frame preserves the standard Pauli commutation relations for $X_i'$, $Y_i'$, and $Z_i'$. Using this notation, the
Hamiltonian can be expressed as

\begin{align}
\begin{split}
H=a_1X_1'+a_2X_2'+\frac{\varepsilon J_1J_2}{a_1a_2}(J_1X_1'+h_1Z_1')(J_2X_2'+h_2Z_2').
\label{eqn:hamiltonianrot}
\end{split}
\end{align}

Using the identity for the exponential of a sum of operators,
\begin{eqnarray}
e^{-it(A+\varepsilon B)}&=&e^{-itA}\left [1+(-it)\varepsilon B-\frac{(-it)^2}{2!}\varepsilon[A,B]\right. \cr
&+&\left.\frac{(-it)^3}{3!}\varepsilon[A,[A,B]]-...\right ]+O(\varepsilon^2),
\label{eqn:operatoridentity}
\end{eqnarray}
with $A=a_1X_1'+a_2X_2'$ and $B=(J_1X_1'+h_1Z_1')(J_2X_2'+h_2Z_2')$, we can evaluate the commutators on the right side and
resum the series, yielding a closed form expression for the evolution operator $e^{-itH}$ to first order in $\varepsilon$.
The result we obtain is
\begin{equation}
e^{-itH}=e^{-it(a_1X_1'+a_2X_2')}\left (1-i\sum_{i'j'}\Delta_{i'j'}^{\text{cap}}\sigma_{i'}\otimes\sigma_{j'}\right ),
\label{eqn:concisekerror}
\end{equation}
where $\sigma_{i'}$ and $\sigma_{j'}$ are the rotated Pauli matrices defined by Eq.\ \eqref{eqn:paulirotation}, and 
the $\Delta_{i'j'}^{\text{cap}}$ are given by
\begin{align}
\Delta_{X_1'X_2'}^{\text{cap}}=\varepsilon&\frac{J_1^2J_2^2}{a_1a_2}t,\nonumber\\
\begin{split}
\Delta_{Y_1'Y_2'}^{\text{cap}}=\varepsilon&\frac{J_1J_2h_1h_2}{2a_1a_2}\Big\{-t\sinc{[2(a_1+a_2)t]}\\&+t\sinc{[2(a_1-a_2)t]}\Big\},
\end{split}\nonumber\\
\begin{split}
\Delta_{Z_1'Z_2'}^{\text{cap}}=\varepsilon&\frac{J_1J_2h_1h_2}{2a_1a_2}\Big\{t\sinc{[2(a_1+a_2)t]}\\&+t\sinc{[2(a_1-a_2)t]}\Big\},
\end{split}\nonumber\\
\Delta_{X_1'Y_2'}^{\text{cap}}=\varepsilon&\frac{J_1^2J_2h_2}{a_1}t^2\sinc^2{(a_2t)},\nonumber\\
\Delta_{X_1'Z_2'}^{\text{cap}}=\varepsilon&\frac{J_1^2J_2h_2}{a_1a_2}t\sinc{(2a_2t)},\nonumber\\
\begin{split}
\Delta_{Y_1'Z_2'}^{\text{cap}}=\varepsilon&\frac{J_1J_2h_1h_2}{2a_1a_2}\Big\{(a_1+a_2)t^2\sinc^2{[(a_1+a_2)t]}\\&+(a_1-a_2)t^2\sinc^2{[(a_1-a_2)t]}\Big\}.
\end{split}
\label{eqn:kerror}
\end{align}
The terms $\Delta_{Y_1'X_2'}^{\text{cap}}$, $\Delta_{Z_1'X_2'}^{\text{cap}}$, and $\Delta_{Z_1'Y_2'}^{\text{cap}}$ can be obtained by swapping the subscripts
1 and 2. We write the error in terms of $\sinc{x}=\lim_{x'\to x}\frac{\sin{x'}}{x'}$ so that the error will be well defined when $a_1-a_2=0$. Since
the rotated Pauli matrices $X_i'$, $Y_i'$, and $Z_i'$ depend on the $J_i$, which changes at different points in time, it is necessary to transform Eq.\
\eqref{eqn:kerror} back to the standard basis using Eq.\ \eqref{eqn:paulirotation}. The result of this substitution, which involves considerable algebra
and is not particularly illuminating, is not shown for the sake of space.

We now consider the addition of charge and field noise on each qubit. If the noise varies slowly compared to the total gate implementation time (i.e., quasistatic
noise, often a very reasonable approximation for semiconductor spin qubits), we can treat the noise as small, unknown shifts to the values $h_i$ and $J_i$. Thus,
the complete Hamiltonian is

\begin{align}
\begin{split}
H&=(h_1+\mathit{dh}_1)X_1+(h_2+\mathit{dh}_2)X_2\\&+(J_1+\mathit{dJ}_1-\varepsilon J_1J_2)Z_1+(J_2+\mathit{dJ}_2-\varepsilon J_1J_2)Z_2\\&+\varepsilon J_1J_2Z_1Z_2.
\label{eqn:hamiltonianwitherr}
\end{split}
\end{align}

We choose to work in the quasistatic noise limit, where these shifts do not directly depend on time. Thus, we will treat $\mathit{dh}_i$ as constant for the entire
pulse sequence, and $\mathit{dJ}_i$ as a function solely of $J_i$. This is generally taken to be a linear relationship due to the underlying dependence of $J_i$ on
the quantum dot detuning\cite{WangPRA2014}; however, since our method is a generalization of \textsc{supcode}, it inherits \textsc{supcode}'s robustness in handling
different dependencies of $\mathit{dJ}_i$ on $J_i$, as long as the forms of such dependencies are known and sufficiently well-behaved. In this work, we assume
$\mathit{dJ}_i=\alpha_iJ_i$ for some small constant parameter $\alpha_i$.  Expansions similar to the one above have already been performed\cite{Kestner2013} for the
case of noise on a single qubit, and the presence of a second qubit will not affect the first order terms of these expansions. Additional terms of order $\varepsilon\textit{dh}_i$
or $\varepsilon\textit{dJ}_i$ will appear, but since both the strength of the coupling and the magnitude of the error are small, such terms are second order and can
be ignored. Combining this with the expansion of the Ising term done above, we obtain
\begin{align}
\begin{split}
e^{-itH}=e^{-it(h_1X_1+J_1Z_1)}e^{-it(h_2X_2+J_2Z_2)}\times\\ \bigg[1-i\sum_i\Delta_i^{\text{q1}}\sigma_i\otimes 1
-i\sum_i\Delta_i^{\text{q2}}1\otimes\sigma_i \\ -i\sum_{ij}\Delta_{ij}^{\text{cap}}\sigma_i\otimes\sigma_j\bigg],
\label{eqn:evolutionoperator}
\end{split}
\end{align}
where the $\Delta_{ij}^{\text{cap}}$ are given by Equation \eqref{eqn:kerror} rotated into the standard basis as discussed above and the $\Delta_i^{\text{q}n}$
are given by
\begin{align}
\begin{split}
\Delta_x^{\text{q}n}=&\frac{2h_n^2a_nt+J_n^2\sin{2a_nt}}{2a_n^3} \mathit{dh}_n\\&+\frac{h_nJ_n(2a_nt-\sin{2a_nt})}{2a_n^3}(\mathit{dJ}_n-\varepsilon J_1J_2),
\end{split}\nonumber\\
\begin{split}
\Delta_y^{\text{q}n}=&\frac{J_n(\cos{2a_nt}-1)}{2a_n^2} \mathit{dh}_n\\&+\frac{h_n(1-\cos{2a_nt})}{2a_n^2}(\mathit{dJ}_n-\varepsilon J_1J_2),
\end{split}\nonumber\\
\begin{split}
\Delta_z^{\text{q}n}=&\frac{h_nJ_n(2a_nt-\sin{2a_nt})}{2a_n^3} \mathit{dh}_n\\&+\frac{2J_n^2a_nt+h_n^2\sin{2a_nt}}{2a_n^3}(\mathit{dJ}_n-\varepsilon J_1J_2).
\end{split}
\label{eqn:singleerror}
\end{align}
There are 27 independent error terms: the 6 components $\Delta_{i}^{\text{q}n}$ each have three terms corresponding to $\mathit{dh}_n$, $\mathit{dJ}_n$,
and $\varepsilon$, plus the 9 components $\Delta_{ij}^{\text{cap}}$ which have only a single term each.

\section{Error Correction}

In order to develop dynamically corrected rotation sequences, it is necessary to examine in general how errors from consecutive rotations are
combined. To that end, let $R$ represent an ideal rotation (with no error or crosstalk) for given values of $J_1$, $J_2$, and $t$; $U$ the
corresponding uncorrected rotation; and $\Delta_R$ the first-order error in the rotation, so that $U=R(1+\Delta_R)$, which has the same form as
Equation \eqref{eqn:evolutionoperator}. We also let $M=R_1R_2\ldots R_m$ be a sequence of $m$ ideal rotations and $\Delta_M$ the total error of the
sequence of corresponding uncorrected rotations, so that $M(1+\Delta_M)=U_1U_2\ldots U_m$. We can then increase the sequence by one additional rotation using
the following equation, which holds to first order in $\Delta_M$ and $\Delta_R$:
\begin{equation}
M(1+\Delta_M)U=MR(1+R^\dagger\Delta_MR +\Delta_R)
\end{equation} 
Recursively applying this equation allows the error of an arbitrary number of gates to be combined. The result is that the total error is a modified sum of the
individual errors $\Delta_{R}$, where each term in the sum is rotated by the inverse of all rotations which occur before (to the right of) $R$ as shown below:
\begin{equation}
\prod_{i=1}^mU_i=\bigg(\prod_{i=1}^mR_i\bigg)\Bigg[1+\sum_{i=1}^m\bigg(\prod_{j=m}^{i+1}R_j^\dagger\bigg)\Delta_{R_i}\bigg(\prod_{j=i+1}^mR_j\bigg)\Bigg]
\label{eqn:errorsum}
\end{equation} 
In order to perform a corrected gate, we follow a strategy similar to the one proposed in Ref.\ \onlinecite{WangPRA2014}, in which a simple set of uncorrected
rotations that implements the desired gate is performed, followed by a longer uncorrected identity operation which is designed to exactly cancel the first order
error in the initial rotations. The error term of the initial rotation, which we will call $\Delta_\text{rot}$, can be directly calculated from Eqs.\
\eqref{eqn:kerror} and \eqref{eqn:singleerror}.  The problem now becomes finding an uncorrected identity with an error term $\Delta_\text{idt}=-\Delta_\text{rot}$,
i.e., an uncorrected identity with a total error that exactly cancels the error in the initial rotation. The simplest way to do this is by introducing
a family of identity operations described by a set of free parameters, writing $\Delta_\text{idt}$ in terms of these parameters, and then numerically solving the
equations $\Delta_\text{idt}+\Delta_\text{rot}=0$. Since there are 27 independent error terms, at least 27 free parameters are required to fully cancel the error.

Again following Ref.\ \onlinecite{WangPRA2014}, we choose to use a family of interrupted identity operations. For a single qubit, such an operation consists of a
$2\pi$ rotation at one value of $J$ interrupted by a $2\pi$ rotation at a different value of $J$.  This second, inner $2\pi$ rotation can again be interrupted, and
so forth for arbitrarily many $2\pi$ rotations. The free parameters in this family of operations include the values of $J$ as well as the angles at which the
interruptions take place.  In this way, a family of identity operations which depends on any number of parameters can be generated.  It is then simply a matter of 
extending this notion to a system of two qubits. We first address the case where $h_1=h_2$, and then discuss the subtleties which arise when $h_1$ and $h_2$
are different.

\subsection{Identity with $h_1=h_2$}

\begin{figure*}
	\includegraphics[width=1.7\columnwidth]{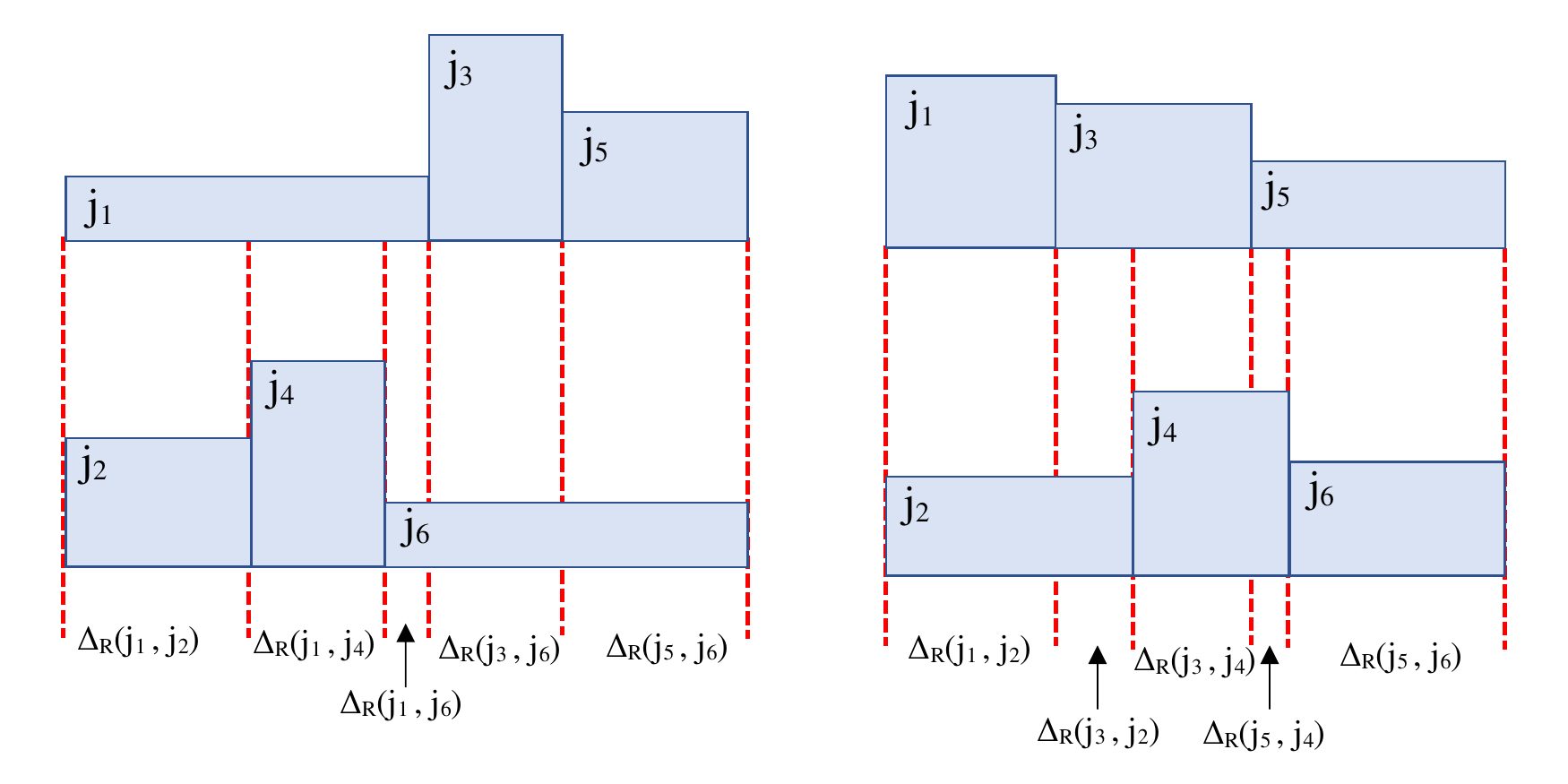}
	\caption{\textbf{Left:} A na\"ive uncorrected identity operation consisting of a sequence of three $2\pi$ rotations on each of qubit 1 (top) and qubit 2 (bottom).
	\textbf{Right:} A similar uncorrected identity, but with different choices of the parameters $j_1 ... j_6$.  As shown, the arguments of the error terms $\Delta_R$
	are different for different values of $j_1 ... j_6$}
	\label{fig:naivepulse}
\end{figure*}

A na\"ive approach to finding an interrupted identity for two qubits would be to use a different sequence of interrupted $2\pi$ rotations on each qubit. There is, however,
one major problem with such a scheme: the time taken to perform a $2\pi$ rotation depends on the value of $J$, so the portion of the first sequence which coincides with a
part of the second sequence is also dependent on the values $J$ takes. Then, for completely arbitrary parameters, it is not possible to find a simple general expression
for the total error. We illustrate this problem with an example shown in Fig.\ \ref{fig:naivepulse}. On each side is shown a pulse consisting of three $2\pi$ rotations
on each qubit. In practice, these would be nested, but we use consecutive $2\pi$ rotations for the sake of this discussion. The total error would be a modified sum of
the $\Delta_R$ terms using Eq.\ \eqref{eqn:errorsum}. The problem is that the arguments of $\Delta_R$ themselves are different for different values of $j_n$. For
example, the term $\Delta_R(j_1,j_6)$ appears on the left of Fig.\ \ref{fig:naivepulse} but not the right, and $\Delta_R(j_3,j_4)$ appears on the right but not the left.
Thus, while this is a perfectly valid uncorrected identity operation, in practice it would be very difficult to tune the parameters $j_n$ to cancel a given error matrix
due to the difficulty of finding a closed-form expression that is valid for all choices of $j_n$.  Theoretically, this is a perfectly allowed dynamical decoupling protocol,
but numerically solving for the specific values of the parameters is challenging, if not impossible.

\begin{figure}
	\includegraphics[width=\columnwidth]{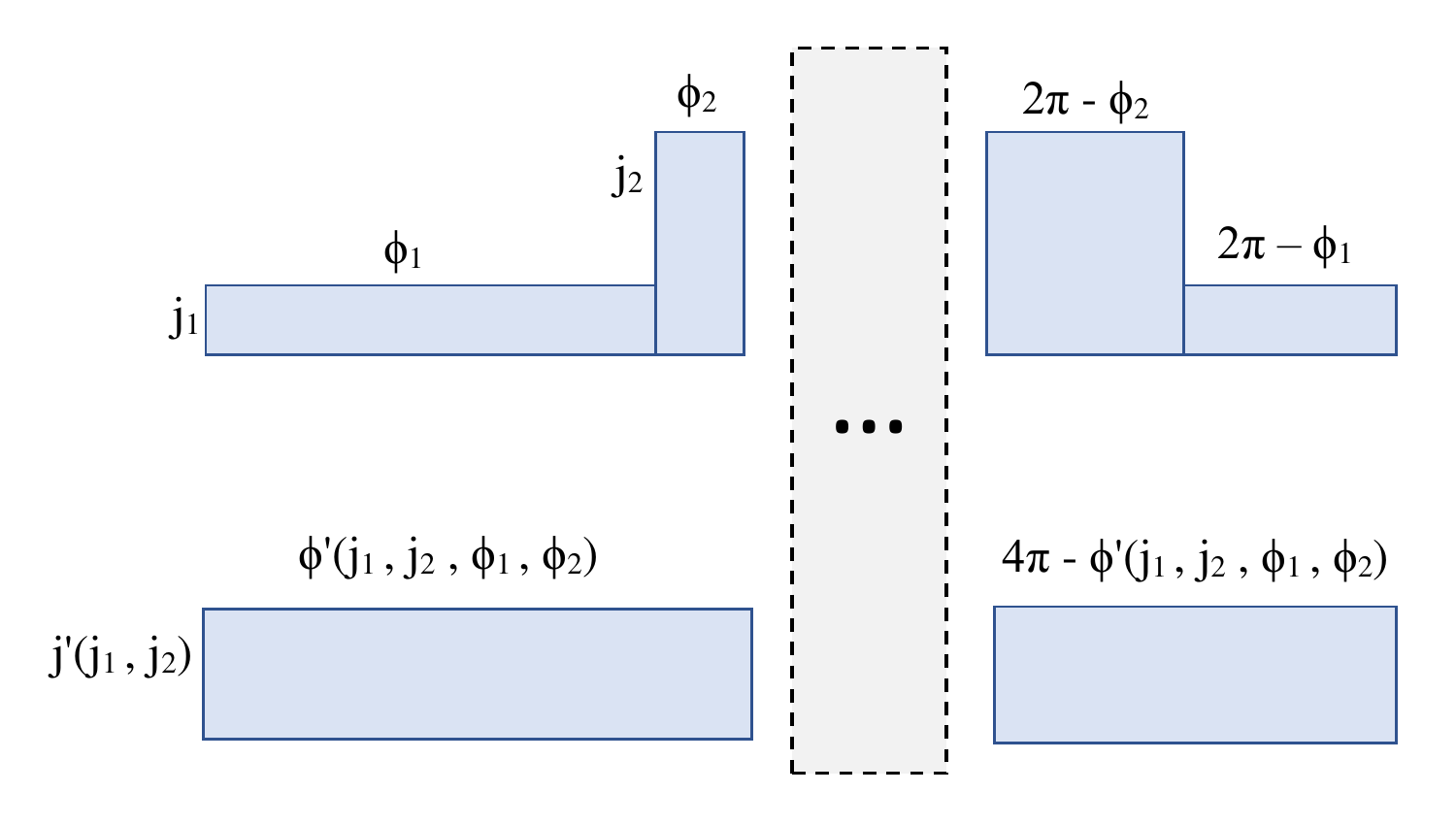}
	\caption{An uncorrected identity for which a closed form expression for arbitrary choices of the parameters is easy to obtain.  There are four free parameters, namely,
	$j_1$, $j_2$, $\phi_1$, and $\phi_2$.  The values of $j'$ and $\phi'$ are determined by the other parameters.}
	\label{fig:betterpulse}
\end{figure}

To circumvent this problem, an identity operation for the entire two-qubit system which has tunable parameters is required. Additionally, we require that it should not be symmetric
between the two qubits, since the errors accumulated on each qubit may differ. The simplest such identity operation consists of a pair of interrupted $2\pi$ rotations
on one qubit, and one $4\pi$ rotation on the other, as shown in Fig.\ \ref{fig:betterpulse}. The value of $j'$ for the $4\pi$ rotation is chosen so that it takes the same
amount of time as the two $2\pi$ rotations. Note that the condition $h_1=h_2$ is implicitly used here, since a value of $j'$ which allows the pulses on each qubit to take the
same time is not guaranteed to exist if $h_1$ and $h_2$ are not equal, as discussed in the next section. This identity operation can then be nested many times, alternating which
qubit undergoes the $4\pi$ rotation and the two $2\pi$ rotations.

We now define $h_1=h_2=h$ to be the common value of $h_1$ and $h_2$.  This situation is in fact typical for two-qubit experiments, with typical values of $h$ being $30\text{ MHz}$
(Ref.\ \onlinecite{ShulmanScience2012}) and $23\text{ MHz}$ (Ref.\ \onlinecite{MartinsPRL2016}).  We introduce the following notation in order to clearly define the nested identity
operation discussed above. Let $U(J_1,J_2,t)$ be an uncorrected rotation at values $J_1$, and $J_2$, for a time $t$.  Also let $t(j,\phi)$ be the time required to perform
a rotation by an angle of $\phi$ about the axis with the given value of $j$.  Then, $t(j,\phi)$ is given as follows:

\begin{equation}
t(j,\phi)=\frac{\phi}{2\sqrt{h^2+j^2}},
\label{eqn:timeofj}
\end{equation}

Since our scheme requires alternating on which qubit the two $2\pi$ rotations are performed, we define the quantity $U_{(n)}(j,j',\phi)$ to encode this alternation.  The notation
corresponds to Fig. \ref{fig:betterpulse}, where $j$ gives the height of a particular $2\pi$ rotation, and $\phi$ gives the angle at which the rotation is interrupted, while $j'$
is the value of the exchange coupling at which the other qubit is held during this rotation.  Whether qubit 1 or 2 corresponds to $j$ or $j'$ is given by the parity of the subscript
$(n)$, so that in terms of $U(J_1,J_2,t)$ described above, $U_{(n)}(j,j',\phi)$ can be written as

\begin{equation}
U_{(n)}(j,j',\phi)=
\begin{cases}
&U(j,j',t(j,\phi))\text{\qquad if $n$ is odd,} \\
&U(j',j,t(j,\phi))\text{\qquad if $n$ is even.}
\end{cases},
\end{equation}

Because the total time taken to perform the two $2\pi$ rotations must equal the time required to perform the $4\pi$ rotation, the value $j'$ at which the $4\pi$ rotation is performed
must depend on the values of $j$ chosen for the two $2\pi$ rotations.  Thus we define $j'_n$ as the value needed in order to fulfill this condition in terms of the values $j_{2n-1}, j_{2n}$,
and find that $j'_n$ is given as follows:

\begin{equation}
j'_n=\sqrt{-1+\Big(\frac{2\pi}{t(j_{2n-1},2\pi)+t(j_{2n},2\pi)}\Big)^2}.
\label{eqn:jprime}
\end{equation}

One level of the identity operation consists of four rotations, as shown in Fig. \ref{fig:betterpulse}.  Nesting $N$ copies of this identity, will then result in a product of $2N$ partial $2\pi$ rotations followed by the product of the $2N$ completions of these rotations in the reverse order.  Thus, in terms of our previous notation, an $N$th level nested uncorrected identity denoted $I^{(N)}$ is given by

\begin{align}
\begin{split}
I^{(N)}=\prod_{n=1}^N\Bigg[&U_{(n)}\Big(j_{2n-1},j'_n,\phi_{2n-1}\Big)U_{(n)}\Big(j_{2n},j'_n,\phi_{2n}\Big)\Bigg]\\
\prod_{n=N}^1\Bigg[&U_{(n)}\Big(j_{2n},j'_n,2\pi-\phi_{2n}\Big)\\&U_{(n)}\Big(j_{2n-1},j'_n,2\pi-\phi_{2n-1}\Big)\Bigg].
\label{eqn:identity}
\end{split}
\end{align}

This method can correct to first order any single qubit gate or product of single qubit gates. As a demonstration of this procedure, we provide pulse sequences for gates of the form
$R\otimes 1$, where $R$ is one of the 24 Clifford gates. To implement this method of deriving dynamical pulse sequences, we first find short uncorrected pulse sequences which implement
the desired gates. Reference \onlinecite{ThrockmortonArXiv} covers this topic in great detail, and we use the equations for implementing arbitrary rotations found there, many of which have
extra degrees of freedom. (We refer to Ref.\ \onlinecite{ThrockmortonArXiv} for the necessary technical details here.) In principle, these degrees of freedom could be added to the list
of free parameters $j_i$, $\phi_i$ in the uncorrected identity operation, possibly resulting in a marginally more optimal pulse sequence. However, these extra parameters are gate dependent,
and so in order to treat all gates identically, we make arbitrary choices for these parameters, giving preference to choices which have low values of $J_1$ or $J_2$, since the strength
of the capacitive coupling is proportional to $J_1$ and $J_2$. During this initial rotation sequence, which is performed on qubit 1, we perform a single uncorrected $2n\pi$ rotation at
constant $J_2$ on the qubit 2, with $J_2$ and $n$ chosen so that the total time of the initial pulse sequence and the $2n\pi$ rotation are the same.

After calculating the error $\Delta_\text{rot}$ using Equations \eqref{eqn:kerror} and \eqref{eqn:singleerror}, we construct an uncorrected identity of the form given in Equation \eqref{eqn:identity}.
Since 27 parameters are needed, at least a level 7 uncorrected identity is required; however, we have used a level 9 identity in order to introduce extra parameters, which helps the numerical
methods used to solve the equations converge more quickly and consistently. In particular, we minimize the norm of the sum $\Delta_\text{rot}+\Delta_\text{idt}$. A local minimization technique
is used which requires initial values for the parameters, and a poor choice of these values can cause the method to converge to a local minimum not equal to zero. We find that choosing $j_i$ to
have magnitudes alternating between large ($j_{2n-1}>h$) and small ($j_{2n}<h$), and choosing $\phi_i$ close to $\pi$ generally give good results. We leave some room for variation, pseudorandomly
generate many sets of values of $j_i$ and $\phi_i$, and use for the set of initial values the one which gives the smallest norm of $\Delta_\text{rot}+\Delta_\text{idt}$.  During this minimization,
we require that $j_\text{min}\leq j_i\leq j_\text{max}$, with $j_\text{min}$ and $j_\text{max}$ equal to $\frac{1}{30}h$ and $30h$, respectively. These values have been chosen to approximate current
experimental capabilities\cite{BarnesPrivComm}. Our method works with tighter constraints, though a higher level identity (longer pulse) may be required to find a solution.

\subsection{Constraints with $h_1\ne h_2$}

We now address the case where $h_1$ and $h_2$ differ by a known amount $\Delta h$ greater than $\mathit{dh}_1$, $\mathit{dh}_2$, so that terms second order in $\Delta h$ cannot be ignored (if
$\Delta h\ll 1$, then we can simply treat $h_1=h_2$ and let $\Delta h$ be absorbed into the $\mathit{dh}_1$ and $\mathit{dh}_2$ error terms).  We can use a similar method as in subsection A
above, with Equations \eqref{eqn:timeofj}--\eqref{eqn:jprime} becoming

\begin{equation}
h_{(n)}=
\begin{cases}
&h_1\text{\qquad if $n$ is odd}\\
&h_2\text{\qquad if $n$ is even}
\end{cases},
\end{equation}

\begin{equation}
t_{(n)}(j,\phi)=\frac{\phi}{2\sqrt{h_{(n)}^2+j^2}},
\label{eqn:th1h2}
\end{equation}

\begin{equation}
U_{(n)}(j,j',\phi)=
\begin{cases}
&U(j,j',t_{(n)}(j,\phi))\text{\qquad if $n$ is odd,} \\
&U(j',j,t_{(n)}(j,\phi))\text{\qquad if $n$ is even,}
\end{cases},
\end{equation}

\begin{equation}
j'_n=\sqrt{-h_{(n+1)}^2+\Big[\frac{2\pi}{t_{(n)}(j_{2n-1},2\pi)+t_{(n)}(j_{2n},2\pi)}\Big]^2}.
\label{eqn:jph1h2}
\end{equation}

Here, parentheses $(n)$ in the subscript denote that only the parity of $n$ is important and not its value.

The difficulty lies in ensuring that all values of $j_i$ and $j'_i$ are bounded between $j_{\text{min}}$ and $j_{\text{max}}$. When $h_1=h_2$, the constraints $j_{\text{min}}\leq j_i\leq j_{\text{max}}$
automatically enforce $j_{\text{min}}\leq j'\leq j_{\text{max}}$; however, when $h_1\ne h_2$, additional constraints are needed. We approach this problem by noticing that Eq.\ (\ref{eqn:jph1h2}) is
symmetric between $j_{2n-1}$ and $j_{2n}$. This initially motivates us to require that the constraints on $j_{2n-1}$ and $j_{2n}$ be identical as well. This simplifies the initial calculations, but we
also address allowing asymmetric constraints between $j_{2n-1}$ and $j_{2n}$. Solving Eqs.\ \eqref{eqn:th1h2} and \eqref{eqn:jph1h2} along with the inequality, $j_{\text{min}}\leq j'\leq j_{\text{max}}$,
we find the following constraints on $j_i$:

\begin{align}
\begin{split}
\sqrt{j_{\text{min}}^2-h_{(n)}^2+h_{(n+1)}^2}\leq j_{2n-1}, j_{2n}\\j_{2n-1}, j_{2n}\leq\sqrt{j_{\text{max}}^2-h_{(n)}^2+h_{(n+1)}^2}
\end{split}
\end{align}

In practice, $j_{\text{max}}\gg h_1,h_2$, so the maximum $\sqrt{j_{\text{max}}^2-h_{(n)}^2+h_{(n+1)}^2} \approx j_{\text{max}}$, and thus can be ignored.  The minimum is also ignored when
$-h_{(n)}^2+h_{(n+1)}^2<0$, since $j_i\geq j_{\text{min}}$ fulfills this inequality also. This occurs for all odd or even $n$ depending on whether $h_1$ or $h_2$ is larger.  Using this additional
constraint, the same process described above is used to find a pulse sequence, though a longer pulse may be necessary due to the tighter constraints.

Since the pulse sequences we find generally consist of values $j_i$ alternating between large and small (due to our choice of starting values during the minimization process), it is natural to consider
allowing $j_{2n}\geq j_{\text{min}}$ and deriving the minimum constraint on $j_{2n-1}$, which yields

\begin{equation}
j_{2n-1}\geq \Bigg[-h_{(n)}^2+\frac{(h_{(n+1)}^2+j_{\text{min}}^2)(h_{(n)}^2+j_{\text{min}}^2)}{\Big(2\sqrt{h_{(n)}^2+j_{\text{min}}^2}-\sqrt{h_{(n+1)}^2+j_{\text{min}}^2}\Big)^2}\Bigg]^{1/2}.
\end{equation}

Assuming $j_{\text{min}}\ll h_1,h_2$, this simplifies to

\begin{equation}
j_{2n-1}\geq \frac{2h_{(n)}^{3/2}\sqrt{h_{(n+1)}-h_{(n)}}}{2h_{(n)}-h_{(n+1)}}.
\end{equation}

This approach is only valid for $h_{(n+1)}<2h_{(n)}$.  For differences greater than this, a different choice of uncorrected identity is needed, which, although possible, is beyond the scope of
the current work.

\subsection{Results}

\begin{figure*}
	\makebox[\textwidth]{\includegraphics[width=2.1\columnwidth]{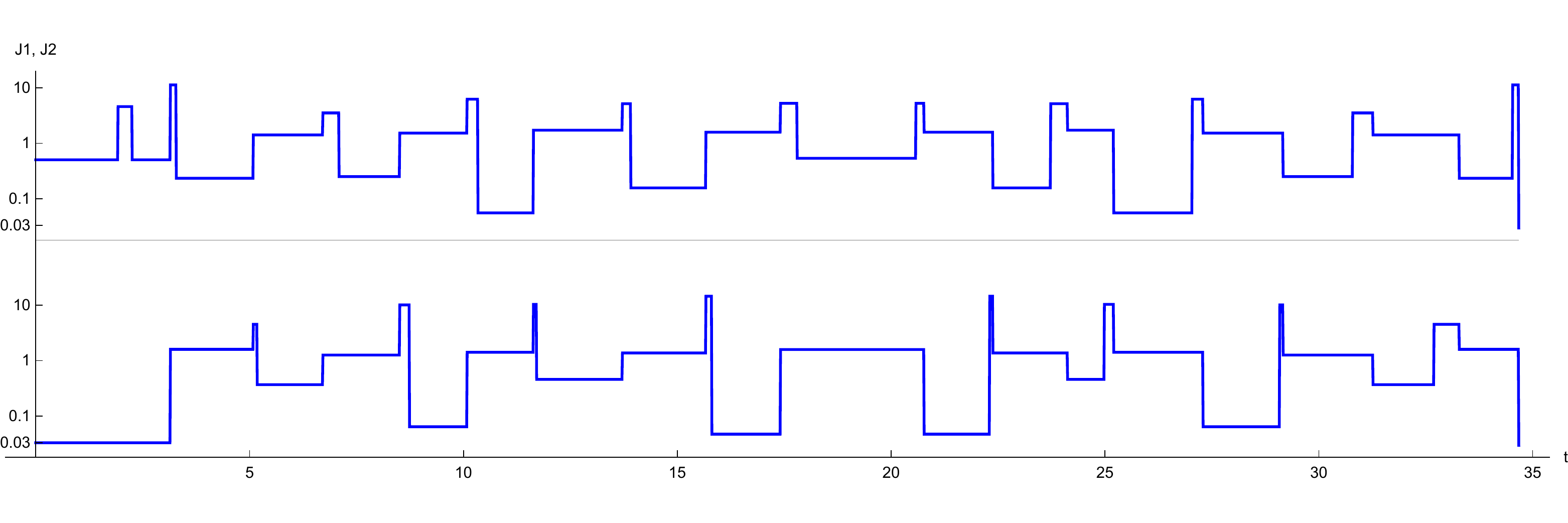}}
	\caption{A corrected pulse implementing a $\pi$ rotation about the $x+y$ axis.  $J_1$ (top) and $J_2$ (bottom) are plotted against time, with $h_1=h_2=1$}
	\label{fig:cxy1log}
\end{figure*}

\begin{figure*}
	\makebox[\textwidth]{\includegraphics[width=1.4\columnwidth]{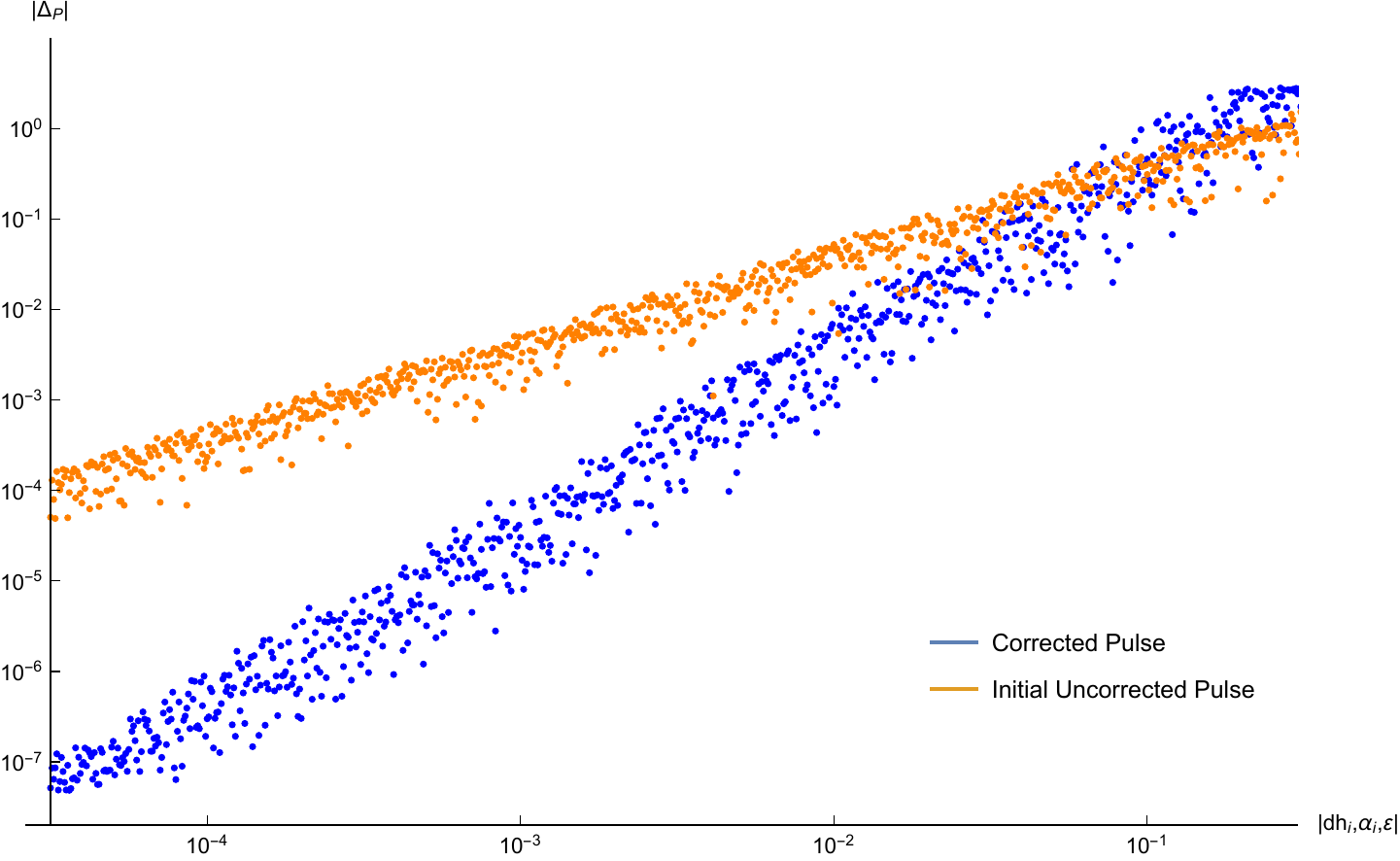}}
	\caption{The total error of corrected (blue) and uncorrected (orange) pulses versus the norm of $\mathit{dh}_i$, $\alpha_i$, $\varepsilon$.  It is clear from the slopes of the lines that the uncorrected
	pulse error is first order and the corrected pulse error is second order in $\mathit{dh}_i$, $\alpha_i$, $\varepsilon$.}
	\label{fig:errorplot}
\end{figure*}

Parameters for the corrected pulse sequences for the 24 Clifford gates for the case where $h_1=h_2=1$ are given in Appendix A.  We have also
considered the simple situation in which, rather than setting $dJ_i=\alpha_i J_i$, with $\alpha_i$ a small constant, we simply let $dJ_i$ be constant for
the duration of the sequence, and provided parameters for the resulting corrected sequences in Appendix B.  The top portions of the tables give
the parameters $j'^{\text{rot}}$, $j_i^{\text{rot}}$, and $t_i^{\text{rot}}$ for the initial uncorrected rotations,
\begin{equation}
U^{\text{rot}} = \prod_iU\Big(j_i^{\text{rot}},j'^{\text{ rot}},t_i^{\text{rot}}\Big),
\label{eqn:initialrot}
\end{equation}
and the bottom portions give the parameters $j_i$ and $\phi_i$ for the uncorrected identity $I^{(9)}$ given by Equation \eqref{eqn:identity}.  As an example,
we plot the error-corrected pulse sequence for the gate $e^{-\frac{i}{2}\frac{X+Y}{\sqrt{2}}\pi}$ in Fig.\ \ref{fig:cxy1log}.

To precisely quantitatively calculate the fidelity of the dynamically corrected gate operations using our pulse sequences starting with given error sets, a detailed randomized benchmarking
analysis is necessary.  Such a randomized benchmarking analysis, which is better done in connection with the actual experimental work implementing our pulse sequences in the ST qubit gate
operations, is well beyond the scope of the current work. However, it is easy to show that the errors in the sequences we found are second order in $\mathit{dh}_i$, $\alpha_i$, and $\varepsilon$,
given that these values are constant, i.e., these sequences correct noise- and crosstalk-induced error to leading order. For a given pulse sequence and a set of values $\mathit{dh}_i$, $\alpha_i$,
and $\varepsilon$, we can evaluate $e^{-itH}$ for each rotation in the pulse sequence using the full Hamiltonian with errors given in Equation \eqref{eqn:hamiltonianwitherr}. By combining these
rotations with error and comparing to the ideal rotation, the total error of the pulse sequence, $\Delta_P(\mathit{dh}_i,\alpha_i,\varepsilon)$, can be determined for the given values of
$\mathit{dh}_i$, $\alpha_i$, and $\varepsilon$. Taking the norm of $\Delta_P$, defined as $|\Delta_P|=\sqrt{\text{Tr }\Delta_P^\dagger\Delta_P}$, we have measure of the error of a given pulse
sequence. For each sequence, we generate many sets of values $\mathit{dh}_i$, $\alpha_i$, and $\varepsilon$, and plot $|\Delta_P|$ against the norm of $\big(\mathit{dh}_1, \mathit{dh}_2, \alpha_1, \alpha_2, \varepsilon\big)$, and find that the total error is clearly second order in $\mathit{dh}_i$, $\alpha_i$, and $\varepsilon$.  We show an example of such a plot, in this case for the gate
$e^{\frac{-i}{2}\frac{X+Y}{\sqrt{2}}\pi}$, in Figure \ref{fig:errorplot}. We generate plots like this for all 24 pulses and find that they are qualitatively identical to the one shown.
As can be seen from Figure \ref{fig:errorplot}, an initial error of $10^{-3}$ can be corrected by two or three orders of magnitude by our pulse sequence whereas the corresponding error correction
for starting errors of $10^{-1}$ is more modest (in the range of one order of magnitude or less).  A full randomized benchmarking analysis is likely to give numbers close to these direct estimates
based on specific pulse sequences.  If the uncorrected system has a fidelity of $95\%$, our dynamical decoupling scheme should be able to improve the fidelity to well above $99\%$, but a starting
fidelity of $90\%$ may be too low for our scheme to improve it above $99\%$.  It is also important to note that although our scheme corrects quite well against correlated errors, if the system is
dominated by white noise or uncorrelated errors, uncorrected pulses are still favorable.  Notably this sets limits on the minimum precision needed to implement the corrected gates.  In particular,
a precision of at least 0.1\% for each of the parameters is needed in  order for the corrected pulses to offer an advantage over the uncorrected gates.

\section{Conclusion} \label{Sec:Conclusion}
We have addressed the issue of crosstalk between two capacitively coupled singlet-triplet qubits.  While such a coupling is necessary to perform multiqubit operations, it also has a detrimental effect
on single-qubit operations---it causes unintentional rotation of the other qubit(s), and also causes errors in the intentional rotation of the qubit under consideration.  As a result, it is crucial to
develop a method for correcting for not only noise-induced error, but for crosstalk-induced error as well.  Our proposed dynamical decoupling scheme does this precisely.  Our method is a generalization
of the \textsc{supcode} technique of Refs.\ \onlinecite{WangNatComm2012}, \onlinecite{WangPRA2014}, and \onlinecite{ThrockmortonArXiv}.  We first perform our na\"ive pulse sequence for implementing a
given single-qubit gate, followed by an uncorrected identity operation with its parameters arranged in such a way as to cancel the effects of noise and crosstalk to leading order.  Unlike in the
single-qubit case already considered in the literature, the case of two capacitively-coupled qubits provides additional challenges.  Because of the fact that the capacitive coupling is proportional to
the exchange couplings of the two qubits, we cannot make these couplings too large, lest we make the capacitive coupling, and thus the crosstalk, large as well.  We must also ensure that the pulse sequences applied to each qubit take the same amount of time to complete.  Our method for implementing the uncorrected identity, in which we divide the identity into segments within which one qubit is subject to
two pulses and the other to just one of identical duration, allows us to address both of these issues.  We provide an analysis of one of the pulse sequences that results from this procedure, namely, that
of a rotation by $\pi$ about the $\hat{\vec{x}}+\hat{\vec{y}}$ axis, and also provide parameters for all $24$ of the Clifford gates used in randomized benchmarking simulations in the Appendix.  We also
show that these pulses, as claimed, do indeed cancel the error in the na\"ive pulse sequences to leading order by plotting the error as a function of the noise and crosstalk strength.  Throughout, we
assume upper and lower bounds on the exchange couplings that approximate experimental constraints.

All of the results that we provide assume that the magnetic field gradient on the two qubits is the same for both.  It is entirely possible for the gradients to differ.  While our sequences can correct
for small variations due to imperfections in fabrication of the micromagnets or polarization of the nuclei used to realize the gradients (we simply add them to the error terms), we do not derive results
for cases in which the difference in the gradient between the two qubits is significant, whether by design or by accident.  We have, however, provided some discussion of this case; we showed that additional
complications arise, in the form of tighter constraints on the allowed exchange coupling strengths than we would expect entirely from experimental constraints.  We also note that, throughout this work, we
assumed that the noise in the system is quasistatic.  While this is often a good approximation\cite{MartinsPRL2016}, the noise in actual experimental systems is known to follow a power-law
spectrum\cite{DialPRL2013,MedfordPRL2012} ($1/f^\alpha$).  We see, however, from the previous work\cite{WangPRA2014,ThrockmortonArXiv} that, even in this scenario, similar approaches to that which we adopt
here to combat the effects of noise still result in noticeable improvement in gate fidelities.  We thus expect similar results for the capacitively-coupled qubits considered here.

We also note that capacitive coupling is not the only type of coupling that can be realized between two singlet-triplet qubits.  Exchange coupling is another type of coupling that can exist; we simply
couple one quantum dot from one qubit to a quantum dot in the other qubit using a Heisenberg exchange interaction.  This can be realized experimentally using a four-quantum-dot device operated as two
singlet-triplet qubits, with the ``middle'' potential barrier used to control the exchange coupling between the qubits.  This offers two advantages, but also a drawback.  First of all, this type of coupling
enables us to control the coupling of the qubits independently of the intra-qubit exchange couplings, unlike capacitive coupling, which depends on the intra-qubit exchange couplings.  Second of all, it is
compatible with the barrier, or symmetric, control scheme demonstrated in Ref.\ \onlinecite{MartinsPRL2016}, which results in an order of magnitude less charge noise than tilt control.  Realizing the
capacitive coupling studied in this work, on the other hand, requires us to tilt the potential profile in order to produce the electric dipole moments needed.  Therefore, we expect that a putative
set of dynamically corrected pulse sequences for implementing single-qubit gates on exchange-coupled singlet-triplet qubits would avoid a number of the issues that arise with capacitively coupled qubits;
we would no longer be required to use tilt control, thus allowing us to mitigate the effects of charge noise, and we would be able to use faster pulses.  However, the disadvantage is that exchange coupling
introduces yet another source of error beyond crosstalk, namely, leakage errors.  It is possible for this coupling to take the qubits out of their logical subspace; for example, it could leave one qubit in
the $\ket{\uparrow\uparrow}$ state and the other in the $\ket{\downarrow\downarrow}$ state.  This would present an additional challenge to overcome in developing dynamically corrected pulse sequences for
such a system.  While a study of error correction in exchange-coupled singlet-triplet qubits would be interesting, it is beyond the scope of this work.  In the end, whether capacitive coupling or exchange coupling is a better avenue for future progress in ST qubits depends quite a bit not only on the details of the noise and crosstalk in the systems, but also on the experimental ability to control the leakage error.  We mention in this context that the capacitively coupled ST qubits are the only semiconductor spin qubit systems to have demonstrated two-qubit gate controlled operations, leading to our decision to
study the ST qubits in depth in this work.

In conclusion, we have developed detailed dynamical decoupling pulse sequences for suppressing crosstalk and noise errors in capacitively coupled singlet-triplet spin qubits and have explicitly demonstrated their efficacy by showing that the corrected sequences manifest orders of magnitude lower errors than the na\"ive uncorrected sequences.

\acknowledgements
The authors thank Jason Kestner for helpful discussions. This work is supported by Laboratory for Physical Sciences.

\appendix

\section{Tables of dynamically corrected pulse sequence parameters}

We present here the parameters for the dynamically-corrected pulse sequences that we have derived. The complete pulses are given by Eqs.\ \eqref{eqn:initialrot} and \eqref{eqn:identity} with parameters given in Tables \ref{Tab:RotParamsI}--\ref{Tab:RotParamsIII}.  All $j$ values are in units of $h$.
\begin{table*}
	\centering
		\begin{tabular}{| c | c | c | c | c | c | c | c | c |}
		\hline
		Axis & Identity & $x$ & $x$ & $x$ & $y$ & $y$ & $y$ & $z$ \\
		\hline
		Angle & \text{---} & $\pi/2$ & $\pi$ & $3\pi/2$ & $\pi/2$ & $\pi$ & $3\pi/2$ & $\pi/2$ \\
		\hline
		$j'^{\text{rot}}$ & 0.03333 & 0.03333 & 0.03333 & 0.03333 & 0.03333 & 0.03333 & 0.03333 & 0.03333 \\
		\hline
		$j_1^{\text{rot}}$ & 0.03333 & 0.03333 & 0.03333 & 0.03333 & 6.43043 & 7.31731 & 8.19725 & 1.00000 \\
		\hline
		$j_2^{\text{rot}}$ & \text{---} & 0.88348 & 1.73498 & 3.88089 & 0.73084 & 0.75954 & 0.78254 & 0.21835 \\
		\hline
    	$j_3^{\text{rot}}$ & \text{---} & 0.03333 & 0.03333 & 0.03333 & 6.43043 & 7.31731 & 8.19725 & 1.00000 \\
		\hline
		$j_4^{\text{rot}}$ & \text{---} & \text{---} & \text{---} & \text{---} & 0.73084 & 0.75954 & 0.78254 & \text{---} \\
		\hline
		$j_5^{\text{rot}}$ & \text{---} & \text{---} & \text{---} & \text{---} & 6.43043 & 7.31731 & 8.19725 & \text{---} \\
		\hline
		$t_1^{\text{rot}}$ & 3.13985 & 0.40660 & 0.79532 & 1.18104 & 0.12069 & 0.10635 & 0.09511 & 0.61342 \\
		\hline
		$t_2^{\text{rot}}$ & \text{---} & 2.32664 & 1.54921 & 0.77777 & 1.26821 & 1.25089 & 1.23705 & 1.91300 \\
		\hline
		$t_3^{\text{rot}}$ & \text{---} & 0.40660 & 0.79532 & 1.18104 & 0.12069 & 0.21269 & 0.28532 & 0.61342 \\
		\hline
		$t_4^{\text{rot}}$ & \text{---} & \text{---} & \text{---} & \text{---} & 1.26821 & 1.25089 & 1.23705 & \text{---} \\
		\hline
		$t_5^{\text{rot}}$ & \text{---} & \text{---} & \text{---} & \text{---} & 0.36206 & 0.31904 & 0.28532 & \text{---} \\
		\hline
		$j_1$ & 4.25455 & 8.64919 & 8.97960 & 11.2324 & 4.69870 & 11.7867 & 16.5123 & 2.69759 \\
		\hline
		$j_2$ & 0.09332 & 0.23284 & 0.05770 & 0.15526 & 0.45052 & 0.04100 & 0.28630 & 0.18032 \\
		\hline
		$j_3$ & 8.02055 & 2.50370 & 10.2303 & 10.1333 & 14.3073 & 19.4429 & 3.79908 & 20.4549 \\
		\hline
		$j_4$ & 0.06985 & 0.39326 & 0.08544 & 0.13815 & 0.06080 & 0.05751 & 0.35525 & 0.07026 \\
		\hline
		$j_5$ & 10.6268 & 4.16386 & 4.11884 & 3.33988 & 7.62621 & 16.7854 & 4.50812 & 12.0166 \\
		\hline
		$j_6$ & 0.21389 & 0.17203 & 0.06055 & 0.07889 & 0.55734 & 0.32611 & 0.04356 & 0.39711 \\
		\hline
		$j_7$ & 7.37015 & 5.72370 & 3.90696 & 2.96801 & 8.28095 & 16.3807 & 9.11133 & 3.51155 \\
		\hline
		$j_8$ & 0.08986 & 0.05960 & 0.26106 & 0.09212 & 0.05088 & 0.03886 & 0.26085 & 0.10698 \\
		\hline
		$j_9$ & 7.64161 & 3.82995 & 2.06453 & 6.86584 & 4.04735 & 25.1775 & 6.51133 & 5.49661 \\
		\hline
		$j_{10}$ & 0.17317 & 0.07476 & 0.08354 & 0.05764 & 0.67310 & 0.06981 & 0.05847 & 0.10551 \\
		\hline
		$j_{11}$ & 4.66221 & 11.8434 & 12.2670 & 7.57279 & 15.3226 & 21.3511 & 5.54262 & 6.12039 \\
		\hline
		$j_{12}$ & 0.13532 & 0.28465 & 0.16320 & 0.30184 & 0.08241 & 0.05312 & 0.56561 & 0.08912 \\
		\hline
	    $j_{13}$ & 6.51752 & 9.38462 & 8.88344 & 3.45868 & 3.78001 & 2.49265 & 5.73037 & 7.88694 \\
		\hline
		$j_{14}$ & 0.06160 & 0.07390 & 0.06824 & 0.17500 & 0.11185 & 0.34648 & 0.06432 & 0.20510 \\
		\hline
		$j_{15}$ & 9.20117 & 6.50713 & 8.40965 & 7.15013 & 6.01842 & 5.86112 & 9.20095 & 11.0451 \\
		\hline
		$j_{16}$ & 0.06356 & 0.09257 & 0.35852 & 0.13261 & 0.31594 & 0.11879 & 0.07149 & 0.07399 \\
		\hline
		$j_{17}$ & 4.56705 & 7.29008 & 9.44018 & 6.36630 & 8.37878 & 15.1228 & 21.0147 & 7.51700 \\
		\hline
		$j_{18}$ & 0.09158 & 0.42999 & 0.11956 & 0.26920 & 0.46082 & 0.31023 & 0.05848 & 0.27828 \\
		\hline
		$\phi_1$ & 2.55322 & 4.66206 & 0.30870 & 4.51951 & 5.52197 & 3.64920 & 2.80854 & 3.21494 \\
		\hline
		$\phi_2$ & 3.00455 & 3.23089 & 2.12926 & 2.05844 & 5.00581 & 2.83173 & 4.02044 & 4.77361 \\
		\hline
		$\phi_3$ & 3.33459 & 4.11162 & 3.92240 & 4.21749 & 4.04230 & 3.18274 & 1.71682 & 2.38512 \\
		\hline
		$\phi_4$ & 2.96052 & 2.88401 & 4.33729 & 3.53703 & 2.73821 & 3.72700 & 3.01457 & 3.57115 \\
		\hline
		$\phi_5$ & 5.09619 & 3.21946 & 1.39096 & 2.21303 & 2.27049 & 0.92820 & 3.71351 & 3.80415 \\
		\hline
		$\phi_6$ & 2.33360 & 3.35709 & 3.42368 & 3.34577 & 4.07169 & 3.89570 & 2.73709 & 2.63703 \\
		\hline
		$\phi_7$ & 5.20802 & 1.84857 & 3.66689 & 4.25246 & 3.76942 & 2.01892 & 4.46601 & 2.88075 \\
		\hline
		$\phi_8$ & 3.71262 & 3.47093 & 2.51816 & 1.64612 & 3.20440 & 2.61221 & 2.86863 & 3.03145 \\
		\hline
		$\phi_9$ & 0.46072 & 3.30053 & 2.09375 & 2.84958 & 1.39429 & 4.71962 & 4.02309 & 3.03890 \\
		\hline
		$\phi_{10}$ & 3.39227 & 3.37554 & 3.27396 & 3.30266 & 3.68709 & 3.88283 & 3.29495 & 2.73878 \\
		\hline
		$\phi_{11}$ & 2.55021 & 5.62423 & 1.71873 & 1.59029 & 1.43437 & 2.62054 & 0.18895 & 5.34732 \\
		\hline
		$\phi_{12}$ & 2.65459 & 1.74187 & 4.45157 & 4.16496 & 4.23973 & 2.57055 & 4.15438 & 3.30636 \\
		\hline
		$\phi_{13}$ & 5.21937 & 4.40566 & 2.93057 & 2.22547 & 2.23942 & 2.75109 & 3.17595 & 5.68995 \\
		\hline
		$\phi_{14}$ & 2.91567 & 2.92878 & 2.85932 & 3.42050 & 2.13964 & 2.91979 & 2.97230 & 1.91518 \\
		\hline
		$\phi_{15}$ & 3.79412 & 2.44601 & 4.29348 & 3.91960 & 4.96730 & 3.87946 & 3.41412 & 4.76460 \\
		\hline
		$\phi_{16}$ & 2.13795 & 2.96169 & 3.57123 & 2.51887 & 2.08244 & 2.82506 & 3.57640 & 2.60517 \\
		\hline
		$\phi_{17}$ & 4.12557 & 2.35265 & 0.43890 & 4.22659 & 5.48983 & 3.28908 & 2.47360 & 2.86459 \\
		\hline
		$\phi_{18}$ & 3.14159 & 3.14159 & 3.14159 & 3.14159 & 3.14159 & 3.14159 & 3.14159 & 3.14159 \\
		\hline
		\end{tabular}
	\caption{Parameters for the dynamically-corrected identity operation, the $x$ and $y$ rotations, and the $z$ rotation
	by $\frac{\pi}{2}$.}
	\label{Tab:RotParamsI}
\end{table*}
\begin{table*}
	\centering
		\begin{tabular}{| c | c | c | c | c | c | c | c | c |}
		\hline
		Axis & $z$ & $z$ & $\hat{\vec{x}}+\hat{\vec{y}}$ & $\hat{\vec{x}}-\hat{\vec{y}}$ & $\hat{\vec{x}}+\hat{\vec{z}}$ & $\hat{\vec{x}}-\hat{\vec{z}}$ & $\hat{\vec{y}}+\hat{\vec{z}}$ & $\hat{\vec{y}}-\hat{\vec{z}}$ \\
		\hline
		Angle & $\pi$ & $3\pi/2$ & $\pi$ & $\pi$ & $\pi$ & $\pi$ & $\pi$ & $\pi$ \\
		\hline
		$j'^{\text{rot}}$ & 0.03333 & 0.03333 & 0.03333 & 0.03333 & 2.64575 & 0.57485 & 0.03333 & 0.03333 \\
		\hline
		$j_1^{\text{rot}}$ & 2.00000 & 3.00000 & 0.50509 & 0.50509 & 1.00000 & 2.53276 & 1.71541 & 1.71541 \\
		\hline
		$j_2^{\text{rot}}$ & 0.45510 & 0.67658 & 4.57172 & 4.57172 & \text{---} & 0.03333 & 0.13012 & 0.13012 \\
		\hline
		$j_3^{\text{rot}}$ & 2.00000 & 3.00000 & 0.50509 & 0.50509 & \text{---} & 2.53276 & 1.71541 & 1.71541 \\
		\hline
		$j_4^{\text{rot}}$ & \text{---} & \text{---} & \text{---} & \text{---} & \text{---} & \text{---} & \text{---} & \text{---} \\
		\hline
		$j_5^{\text{rot}}$ & \text{---} & \text{---} & \text{---} & \text{---} & \text{---} & \text{---} & \text{---} & \text{---} \\
		\hline
		$t_1^{\text{rot}}$ & 0.50027 & 0.42475 & 1.91411 & 0.89008 & 1.11072 & 0.57686 & 0.51303 & 1.06916 \\
		\hline
		$t_2^{\text{rot}}$ & 2.13931 & 2.29036 & 0.33565 & 0.33565 & \text{---} & 1.56992 & 1.55767 & 1.55767 \\
		\hline
		$t_3^{\text{rot}}$ & 0.50027 & 0.42475 & 0.89008 & 1.91411 & \text{---} & 0.57686 & 1.06916 & 0.51303 \\
		\hline
		$t_4^{\text{rot}}$ & \text{---} & \text{---} & \text{---} & \text{---} & \text{---} & \text{---} & \text{---} & \text{---} \\
		\hline
		$t_5^{\text{rot}}$ & \text{---} & \text{---} & \text{---} & \text{---} & \text{---} & \text{---} & \text{---} & \text{---} \\
		\hline
		$j_1$ & 14.9279 & 23.9766 & 11.2197 & 5.88326 & 0.96910 & 4.21145 & 7.89663 & 10.5977 \\
		\hline
		$j_2$ & 0.21741 & 0.08751 & 0.23548 & 0.06925 & 0.33248 & 0.04880 & 0.16242 & 0.15088 \\
		\hline
		$j_3$ & 2.80327 & 12.5873 & 4.50703 & 6.85985 & 10.5444 & 3.73761 & 9.24311 & 4.57720 \\
		\hline
		$j_4$ & 0.24971 & 0.29095 & 0.36990 & 0.19478 & 0.08280 & 0.20116 & 0.05004 & 0.14168 \\
		\hline
		$j_5$ & 13.6280 & 5.05654 & 3.51726 & 3.10676 & 8.22643 & 6.15100 & 11.9739 & 11.9299 \\
		\hline
		$j_6$ & 0.11850 & 0.12464 & 0.25159 & 0.12773 & 0.41291 & 0.07436 & 0.37828 & 0.04724 \\
		\hline
		$j_7$ & 5.43618 & 10.4596 & 10.0448 & 6.14212 & 10.9543 & 6.86964 & 3.16910 & 11.1235 \\
		\hline
		$j_8$ & 0.19053 & 0.16828 & 0.06464 & 0.10770 & 0.05153 & 0.10872 & 0.09210 & 0.19887 \\
		\hline
		$j_9$ & 2.93129 & 25.2507 & 6.21325 & 3.62073 & 3.99054 & 5.02616 & 5.40783 & 8.30899 \\
		\hline
		$j_{10}$ & 0.17401 & 0.24332 & 0.05614 & 0.04634 & 0.56279 & 0.30939 & 0.24328 & 0.23214 \\
		\hline
		$j_{11}$ & 5.34895 & 13.9661 & 10.2701 & 7.06335 & 1.16160 & 19.2500 & 12.1845 & 10.3575 \\
		\hline
		$j_{12}$ & 0.40949 & 0.45280 & 0.45991 & 0.30292 & 0.04781 & 0.12683 & 0.08331 & 0.32868 \\
		\hline
		$j_{13}$ & 22.1116 & 19.8886 & 5.15864 & 5.85519 & 5.87043 & 19.8651 & 3.88747 & 9.48442 \\
		\hline
		$j_{14}$ & 0.04834 & 0.04964 & 0.15747 & 0.13325 & 0.21666 & 0.06943 & 0.15774 & 0.07721 \\
		\hline
		$j_{15}$ & 8.28348 & 20.8913 & 14.3910 & 4.07973 & 7.49208 & 7.38150 & 14.5753 & 6.62029 \\
		\hline
		$j_{16}$ & 0.03776 & 0.08864 & 0.04757 & 0.06969 & 0.04560 & 0.04795 & 0.08895 & 0.08536 \\
		\hline
		$j_{17}$ & 17.3790 & 11.8760 & 5.23171 & 7.38525 & 11.1053 & 9.76455 & 5.52560 & 9.49243 \\
		\hline
		$j_{18}$ & 0.28731 & 0.13834 & 0.53640 & 0.31872 & 0.06706 & 0.20667 & 0.56000 & 0.16461 \\
		\hline
		$\phi_1$ & 0.17390 & 4.46173 & 3.23390 & 2.22917 & 3.39790 & 5.28562 & 0.65047 & 0.63295 \\
		\hline
		$\phi_2$ & 3.09168 & 2.33997 & 3.70695 & 3.85984 & 3.95766 & 3.62949 & 2.92893 & 3.60713 \\
		\hline
		$\phi_3$ & 3.07075 & 4.39117 & 0.80794 & 2.98899 & 4.26243 & 1.80037 & 3.38641 & 2.60625 \\
		\hline
		$\phi_4$ & 3.39482 & 3.60846 & 3.26099 & 2.19380 & 3.35828 & 3.33234 & 2.88631 & 2.80943 \\
		\hline
		$\phi_5$ & 2.93283 & 2.65185 & 2.77504 & 4.15096 & 1.54770 & 2.81655 & 2.51553 & 4.12300 \\
		\hline
		$\phi_6$ & 2.65427 & 3.56274 & 2.93110 & 3.34935 & 3.91568 & 2.96830 & 3.16500 & 2.66819 \\
		\hline
		$\phi_7$ & 4.87590 & 0.86494 & 4.65637 & 1.82555 & 1.90529 & 3.86311 & 4.43373 & 5.59828 \\
		\hline
		$\phi_8$ & 2.51402 & 3.42610 & 2.69527 & 3.99554 & 3.59852 & 3.47884 & 2.45794 & 2.53965 \\
		\hline
		$\phi_9$ & 4.14460 & 2.17747 & 3.14723 & 3.55778 & 2.34369 & 2.19616 & 5.64073 & 4.45949 \\
		\hline
		$\phi_{10}$ & 3.45194 & 3.09249 & 2.60268 & 3.09779 & 4.24466 & 3.00242 & 3.21892 & 3.53377 \\
		\hline
		$\phi_{11}$ & 0.26740 & 5.38412 & 1.58592 & 5.06376 & 0.79352 & 6.05802 & 3.55833 & 0.49920 \\
		\hline
		$\phi_{12}$ & 3.84950 & 2.68984 & 4.40648 & 2.19840 & 2.86710 & 3.79569 & 2.95205 & 3.42713 \\
		\hline
		$\phi_{13}$ & 2.66076 & 3.07483 & 2.14367 & 3.92841 & 5.36926 & 2.38004 & 2.08220 & 2.48413 \\
		\hline
		$\phi_{14}$ & 3.27057 & 3.42195 & 3.54935 & 3.08201 & 1.77750 & 3.01509 & 3.04083 & 3.12396 \\
		\hline
		$\phi_{15}$ & 2.94726 & 2.01300 & 3.96286 & 2.65718 & 4.68514 & 3.74243 & 3.17084 & 3.63762 \\
		\hline
		$\phi_{16}$ & 3.86814 & 2.82538 & 3.21533 & 2.96932 & 2.63704 & 3.38304 & 3.75684 & 3.40975 \\
		\hline
		$\phi_{17}$ & 3.27372 & 3.39624 & 4.21481 & 2.24716 & 3.60030 & 4.05433 & 3.12939 & 3.00546 \\
		\hline
		$\phi_{18}$ & 3.14159 & 3.14159 & 3.14159 & 3.14159 & 3.14159 & 3.14159 & 3.14159 & 3.14159 \\
		\hline
		\end{tabular}
	\caption{Parameters for the dynamically-corrected $z$ rotations	by $\pi$ and $3\pi/2$, and for the rotations by
	$\hat{\vec{x}}\pm\hat{\vec{y}}$, $\hat{\vec{x}}\pm\hat{\vec{z}}$, and $\hat{\vec{y}}\pm\hat{\vec{z}}$.}
	\label{Tab:RotParamsII}
\end{table*}
\begin{table*}
	\centering
		\begin{tabular}{| c | c | c | c | c | c | c | c | c |}
		\hline
			Axis & $\hat{\vec{x}}+\hat{\vec{y}}+\hat{\vec{z}}$ & $\hat{\vec{x}}+\hat{\vec{y}}+\hat{\vec{z}}$ & $-\hat{\vec{x}}+\hat{\vec{y}}+\hat{\vec{z}}$ & $-\hat{\vec{x}}+\hat{\vec{y}}+\hat{\vec{z}}$ & $\hat{\vec{x}}-\hat{\vec{y}}+\hat{\vec{z}}$ & $\hat{\vec{x}}-\hat{\vec{y}}+\hat{\vec{z}}$ & $\hat{\vec{x}}+\hat{\vec{y}}-\hat{\vec{z}}$ & $\hat{\vec{x}}+\hat{\vec{y}}-\hat{\vec{z}}$ \\
		\hline
			Angle & $2\pi/3$ & $4\pi/3$ & $2\pi/3$ & $4\pi/3$ & $2\pi/3$ & $4\pi/3$ & $2\pi/3$ & $4\pi/3$ \\
		\hline
		$j'^{\text{rot}}$ & 0.03333 & 0.03333 & 0.03333 & 0.03333 & 0.03333 & 0.03333 & 0.03333 & 0.03333 \\
		\hline
		$j_1^{\text{rot}}$ & 0.56170 & 0.74456 & 0.32145 & 0.29210 & 0.56170 & 0.74456 & 0.29210 & 0.32145 \\
		\hline
		$j_2^{\text{rot}}$ & 2.41400 & 3.22657 & 14.0228 & 8.36716 & 2.41400 & 3.22657 & 8.36716 & 14.0228 \\
		\hline
		$j_3^{\text{rot}}$ & 0.56170 & 0.74456 & 0.32145 & 0.29210 & 0.56170 & 0.74456 & 0.29210 & 0.32145 \\
		\hline
		$j_4^{\text{rot}}$ & \text{---} & \text{---} & \text{---} & \text{---} & \text{---} & \text{---} & \text{---} & \text{---} \\
		\hline
		$j_5^{\text{rot}}$ & \text{---} & \text{---} & \text{---} & \text{---} & \text{---} & \text{---} & \text{---} & \text{---} \\
		\hline
		$t_1^{\text{rot}}$ & 2.21343 & 1.97092 & 1.17574 & 1.18212 & 0.52564 & 0.54891 & 1.83346 & 1.81513 \\
		\hline
		$t_2^{\text{rot}}$ & 0.40078 & 0.62001 & 0.14898 & 0.12427 & 0.40078 & 0.62001 & 0.12427 & 0.14898 \\
		\hline
		$t_3^{\text{rot}}$ & 0.52564 & 0.54891 & 1.81513 & 1.83346 & 2.21343 & 1.97092 & 1.18212 & 1.17574 \\
		\hline
		$t_4^{\text{rot}}$ & \text{---} & \text{---} & \text{---} & \text{---} & \text{---} & \text{---} & \text{---} & \text{---} \\
		\hline
		$t_5^{\text{rot}}$ & \text{---} & \text{---} & \text{---} & \text{---} & \text{---} & \text{---} & \text{---} & \text{---} \\
		\hline
		$j_1$ & 3.53560 & 7.40870 & 9.30530 & 6.38544 & 10.7248 & 17.2594 & 7.05950 & 15.1935 \\
		\hline
		$j_2$ & 0.23139 & 0.10157 & 0.11817 & 0.12981 & 0.21240 & 0.17560 & 0.32670 & 0.11769 \\
		\hline
		$j_3$ & 16.0966 & 6.97001 & 10.6390 & 6.51320 & 3.11017 & 27.1973 & 1.29162 & 9.17358 \\
		\hline
		$j_4$ & 0.07111 & 0.11694 & 0.08366 & 0.18193 & 0.14994 & 0.03927 & 0.32813 & 0.11144 \\
		\hline
		$j_5$ & 19.3300 & 4.58619 & 6.75684 & 5.18100 & 18.7600 & 5.91114 & 3.97861 & 8.94809 \\
		\hline
		$j_6$ & 0.16332 & 0.06442 & 0.21631 & 0.12154 & 0.07568 & 0.37476 & 0.13895 & 0.33521 \\
		\hline
		$j_7$ & 19.2548 & 5.60181 & 8.66110 & 4.18403 & 12.5498 & 7.98067 & 5.94619 & 12.1638 \\
		\hline
		$j_8$ & 0.04468 & 0.17014 & 0.14226 & 0.18787 & 0.29095 & 0.03475 & 0.05824 & 0.13070 \\
		\hline
		$j_9$ & 13.7738 & 13.4672 & 18.4340 & 3.59364 & 6.79908 & 11.9460 & 3.70590 & 19.3022 \\
		\hline
		$j_{10}$ & 0.08073 & 0.13413 & 0.24654 & 0.04273 & 0.15602 & 0.31084 & 0.04433 & 0.20006 \\
		\hline
		$j_{11}$ & 5.95821 & 7.28592 & 10.7813 & 6.93732 & 6.81217 & 8.42940 & 8.91232 & 6.86166 \\
		\hline
		$j_{12}$ & 0.13448 & 0.40891 & 0.15650 & 0.33408 & 0.31731 & 0.05540 & 0.33998 & 0.06192 \\
		\hline
		$j_{13}$ & 11.2806 & 7.39708 & 7.60260 & 4.74451 & 8.48105 & 21.8867 & 3.84464 & 5.46750 \\
		\hline
		$j_{14}$ & 0.16682 & 0.05707 & 0.30270 & 0.05037 & 0.09169 & 0.16977 & 0.05388 & 0.27407 \\
		\hline
		$j_{15}$ & 21.0041 & 11.2817 & 15.3315 & 5.09639 & 4.75689 & 12.7137 & 6.22348 & 4.54025 \\
		\hline
		$j_{16}$ & 0.12130 & 0.06069 & 0.05541 & 0.04628 & 0.08300 & 0.37435 & 0.08023 & 0.06762 \\
		\hline
		$j_{17}$ & 14.7728 & 8.14723 & 7.37956 & 6.53512 & 11.0940 & 3.36556 & 7.16938 & 22.1340 \\
		\hline
		$j_{18}$ & 0.17654 & 0.17674 & 0.24785 & 0.21341 & 0.13782 & 0.05163 & 0.51527 & 0.21191 \\
		\hline
		$\phi_1$ & 2.53825 & 4.38397 & 2.65058 & 2.65215 & 2.00758 & 2.68429 & 5.33741 & 3.85920 \\
		\hline
		$\phi_2$ & 4.53270 & 2.20273 & 2.38158 & 2.35687 & 3.52438 & 2.75927 & 3.86201 & 3.37853 \\
		\hline
		$\phi_3$ & 2.59203 & 4.26832 & 4.80735 & 3.92194 & 2.59147 & 3.31991 & 2.81744 & 2.98401 \\
		\hline
		$\phi_4$ & 3.75470 & 3.80220 & 3.24547 & 3.85075 & 3.15874 & 3.39732 & 2.94563 & 3.31101 \\
		\hline
		$\phi_5$ & 0.52120 & 1.79639 & 2.94582 & 2.19146 & 3.34913 & 2.54966 & 3.55481 & 5.69859 \\
		\hline
		$\phi_6$ & 2.78943 & 3.64199 & 2.46653 & 2.92552 & 2.74580 & 3.66193 & 3.26094 & 3.42371 \\
		\hline
		$\phi_7$ & 2.39957 & 0.94526 & 3.64461 & 4.13605 & 5.08724 & 4.01620 & 1.84110 & 3.02494 \\
		\hline
		$\phi_8$ & 3.27728 & 3.49039 & 3.53042 & 2.31790 & 2.53522 & 2.67545 & 3.44628 & 2.23430 \\
		\hline
		$\phi_9$ & 4.13341 & 2.34697 & 1.92427 & 3.04953 & 4.33241 & 5.72953 & 3.24584 & 5.21972 \\
		\hline
		$\phi_{10}$ & 3.64683 & 2.75723 & 2.99155 & 3.37997 & 3.63580 & 2.60940 & 3.40564 & 1.96004 \\
		\hline
		$\phi_{11}$ & 4.64988 & 5.83925 & 5.29808 & 0.84137 & 0.28982 & 4.63477 & 5.17615 & 3.73298 \\
		\hline
		$\phi_{12}$ & 2.94845 & 2.76326 & 3.73889 & 4.18084 & 3.55508 & 2.69058 & 1.48630 & 2.67019 \\
		\hline
		$\phi_{13}$ & 6.10583 & 3.74351 & 3.17347 & 1.99631 & 2.54828 & 3.63376 & 3.93282 & 4.10101 \\
		\hline
		$\phi_{14}$ & 3.47632 & 3.35216 & 2.89166 & 3.62070 & 3.25964 & 2.74220 & 3.41431 & 3.90467 \\
		\hline
		$\phi_{15}$ & 1.43617 & 2.16943 & 3.31085 & 2.91108 & 2.93229 & 3.16477 & 1.31813 & 2.69967 \\
		\hline
		$\phi_{16}$ & 3.36972 & 3.06705 & 3.20338 & 3.46981 & 3.72947 & 4.28321 & 3.27278 & 3.30133 \\
		\hline
		$\phi_{17}$ & 3.62018 & 3.37820 & 2.73933 & 4.06683 & 3.07756 & 0.25141 & 2.27238 & 5.24267 \\
		\hline
		$\phi_{18}$ & 3.14159 & 3.14159 & 3.14159 & 3.14159 & 3.14159 & 3.14159 & 3.14159 & 3.14159 \\
		\hline
		\end{tabular}
	\caption{Parameters for the dynamically-corrected $\hat{\vec{x}}\pm\hat{\vec{y}}\pm\hat{\vec{z}}$ rotations.}
	\label{Tab:RotParamsIII}
\end{table*}

\section{Dynamically Corrected Pulse Parameters for Constant $\mathit{dJ_i}$}

Throughout this work, in the main text, we have used a model where $\mathit{dJ_i}$ is proportional to $J_i$. However, other models may exist, and the numerics of our method are able to
generate pulse sequences for a range of different models. To demonstrate this, we also derived pulse sequences for the case where $\mathit{dJ_i}$ is independent of $J_i$, and found that
these sequences also corrected first-order error in the model chosen. The parameters for these sequences are given in Tables \ref{Tab:constantParamsI}--\ref{Tab:constantParamsIII}. In
principle, if experiments warrant generating dynamical decoupling pulse sequences for other possible coupling models for ST qubits, our method can easily be generalized to handle such
scenarios.  As in the previous set of tables, all $j$ values are in units of $h$.

\begin{table*}
	\centering
	\begin{tabular}{| c | c | c | c | c | c | c | c | c |}
		\hline
		Axis & (Identity) & $x$ & $x$ & $x$ & $y$ & $y$ & $y$ & $z$ \\
		\hline
		Angle & --- & $\pi/2$ & $\pi$ & $3\pi/2$ & $\pi/2$ & $\pi$ & $3\pi/2$ & $\pi/2$ \\
		\hline
		$j'^{\text{rot}}$ & $0.03333$ & $0.03333$ & $0.03333$ & $0.03333$ & $0.03333$ & $0.03333$ & $0.03333$ & $0.03333$ \\
		\hline
		$j_1^{\text{rot}}$ & $0.03333$ & $0.03333$ & $0.03333$ & $0.03333$ & $6.43043$ & $7.31731$ & $8.19725$ & $1.00000$ \\
		\hline
		$j_2^{\text{rot}}$ & --- & $0.88348$ & $1.73498$ & $3.88089$ & $0.73084$ & $0.75954$ & $0.78254$ & $0.21835$ \\
		\hline
		$j_3^{\text{rot}}$ & --- & $0.03333$ & $0.03333$ & $0.03333$ & $6.43043$ & $7.31731$ & $8.19725$ & $1.00000$ \\
		\hline
		$j_4^{\text{rot}}$ & --- & --- & --- & --- & $0.73084$ & $0.75954$ & $0.78254$ & --- \\
		\hline
		$j_5^{\text{rot}}$ & --- & --- & --- & --- & $6.43043$ & $7.31731$ & $8.19725$ & --- \\
		\hline
		$t_1^{\text{rot}}$ & $3.13985$ & $0.40660$ & $0.79532$ & $1.18104$ & $0.12069$ & $0.10635$ & $0.09511$ & $0.61342$ \\
		\hline
		$t_2^{\text{rot}}$ & --- & $2.32664$ & $1.54921$ & $0.77777$ & $1.26821$ & $1.25089$ & $1.23705$ & $1.91300$ \\
		\hline
		$t_3^{\text{rot}}$ & --- & $0.40660$ & $0.79532$ & $1.18104$ & $0.12069$ & $0.21269$ & $0.28532$ & $0.61342$ \\
		\hline
		$t_4^{\text{rot}}$ & --- & --- & --- & --- & $1.26821$ & $1.25089$ & $1.23705$ & --- \\
		\hline
		$t_5^{\text{rot}}$ & --- & --- & --- & --- & $0.36206$ & $0.31904$ & $0.28532$ & --- \\
		\hline
		$j_1$ & $5.46325$ & $4.79267$ & $10.3137$ & $19.4597$ & $7.27643$ & $14.6880$ & $12.1954$ & $18.8080$ \\
		\hline
		$j_2$ & $0.11660$ & $0.34036$ & $0.30761$ & $0.09602$ & $0.17189$ & $0.05126$ & $0.17555$ & $0.07608$ \\
		\hline
		$j_3$ & $7.83045$ & $9.43393$ & $15.1785$ & $3.35795$ & $12.5444$ & $4.86269$ & $21.7404$ & $1.77304$ \\
		\hline
		$j_4$ & $0.19007$ & $0.05276$ & $0.06575$ & $0.04726$ & $0.03873$ & $0.06463$ & $0.03914$ & $0.15544$ \\
		\hline
		$j_5$ & $14.0880$ & $8.39676$ & $4.44094$ & $12.2696$ & $14.3122$ & $20.8924$ & $20.9836$ & $11.8583$ \\
		\hline
		$j_6$ & $0.08858$ & $0.55349$ & $0.09376$ & $0.06754$ & $0.05058$ & $0.29920$ & $0.18160$ & $0.08084$ \\
		\hline
		$j_7$ & $7.86256$ & $7.36736$ & $17.6563$ & $4.50145$ & $5.50740$ & $4.24082$ & $3.78485$ & $5.78212$ \\
		\hline
		$j_8$ & $0.24579$ & $0.05809$ & $0.11776$ & $0.20283$ & $0.30428$ & $0.06582$ & $0.06840$ & $0.09301$ \\
		\hline
		$j_9$ & $8.75655$ & $4.55834$ & $23.1891$ & $6.27128$ & $12.4862$ & $6.52354$ & $2.56912$ & $5.09631$ \\
		\hline
		$j_{10}$ & $0.47191$ & $0.48569$ & $0.06621$ & $0.21726$ & $0.06858$ & $0.44181$ & $0.09971$ & $0.33394$ \\
		\hline
		$j_{11}$ & $13.0447$ & $8.70805$ & $6.58776$ & $14.2591$ & $10.3155$ & $11.3152$ & $22.5411$ & $5.15544$ \\
		\hline
		$j_{12}$ & $0.04996$ & $0.04185$ & $0.37826$ & $0.24168$ & $0.10836$ & $0.23690$ & $0.09658$ & $0.33536$ \\
		\hline
		$j_{13}$ & $8.16027$ & $3.65005$ & $11.7940$ & $5.28220$ & $9.68209$ & $15.1274$ & $10.0745$ & $2.95355$ \\
		\hline
		$j_{14}$ & $0.37463$ & $0.45310$ & $0.19711$ & $0.12640$ & $0.06406$ & $0.12887$ & $0.05410$ & $0.07845$ \\
		\hline
		$j_{15}$ & $8.59080$ & $3.72874$ & $14.2332$ & $21.4867$ & $9.12064$ & $23.7044$ & $20.3803$ & $4.59443$ \\
		\hline
		$j_{16}$ & $0.05091$ & $0.09174$ & $0.11715$ & $0.07025$ & $0.05448$ & $0.05521$ & $0.03650$ & $0.06872$ \\
		\hline
		$j_{17}$ & $8.68563$ & $3.72280$ & $19.1950$ & $13.6475$ & $1.30298$ & $18.1969$ & $10.1510$ & $9.98698$ \\
		\hline
		$j_{18}$ & $0.54697$ & $0.74161$ & $0.14063$ & $0.08958$ & $1.17119$ & $0.55345$ & $0.58051$ & $0.44689$ \\
		\hline
		$\phi_1$ & $3.77050$ & $2.08723$ & $4.87126$ & $2.05446$ & $3.66034$ & $2.28203$ & $1.37456$ & $3.35636$ \\
		\hline
		$\phi_2$ & $3.08246$ & $2.19136$ & $2.24372$ & $2.08569$ & $2.01941$ & $2.62145$ & $3.27109$ & $2.90279$ \\
		\hline
		$\phi_3$ & $3.04031$ & $3.88529$ & $4.54342$ & $4.60125$ & $3.91929$ & $4.65150$ & $0.50256$ & $3.47771$ \\
		\hline
		$\phi_4$ & $3.59030$ & $3.15322$ & $3.25836$ & $3.34866$ & $2.82642$ & $3.34064$ & $2.86164$ & $2.84707$ \\
		\hline
		$\phi_5$ & $0.55135$ & $2.71164$ & $2.55084$ & $2.63841$ & $2.92500$ & $1.62537$ & $2.72019$ & $3.38188$ \\
		\hline
		$\phi_6$ & $2.17272$ & $2.90728$ & $2.89031$ & $2.54085$ & $2.96817$ & $3.28533$ & $3.04122$ & $3.85092$ \\
		\hline
		$\phi_7$ & $4.89381$ & $3.66293$ & $2.01301$ & $4.18211$ & $1.81668$ & $3.89362$ & $3.95251$ & $1.10297$ \\
		\hline
		$\phi_8$ & $3.12411$ & $2.66501$ & $4.03373$ & $2.10076$ & $3.85824$ & $2.42106$ & $3.14859$ & $3.67710$ \\
		\hline
		$\phi_9$ & $3.16856$ & $5.35428$ & $0.91936$ & $5.19087$ & $1.22979$ & $4.52306$ & $4.46065$ & $1.64572$ \\
		\hline
		$\phi_{10}$ & $4.44070$ & $2.98250$ & $3.07860$ & $2.49427$ & $3.55619$ & $1.78678$ & $3.99485$ & $3.08432$ \\
		\hline
		$\phi_{11}$ & $3.43482$ & $3.84937$ & $5.27382$ & $6.04607$ & $4.11185$ & $3.63263$ & $1.15521$ & $5.93853$ \\
		\hline
		$\phi_{12}$ & $3.34802$ & $2.63375$ & $3.08498$ & $2.94918$ & $3.42171$ & $2.58892$ & $3.27690$ & $2.30308$ \\
		\hline
		$\phi_{13}$ & $1.45811$ & $2.85588$ & $3.78840$ & $5.01766$ & $3.12784$ & $5.35763$ & $1.04860$ & $3.77584$ \\
		\hline
		$\phi_{14}$ & $3.17489$ & $3.53121$ & $2.87810$ & $2.81372$ & $3.60913$ & $3.02506$ & $2.95315$ & $3.03211$ \\
		\hline
		$\phi_{15}$ & $3.22668$ & $3.15860$ & $2.59690$ & $3.27866$ & $0.85238$ & $4.77421$ & $2.49322$ & $2.77818$ \\
		\hline
		$\phi_{16}$ & $3.22662$ & $3.09129$ & $3.73021$ & $2.69768$ & $3.44236$ & $2.88387$ & $3.38333$ & $2.69569$ \\
		\hline
		$\phi_{17}$ & $4.00536$ & $4.29243$ & $1.89763$ & $2.89193$ & $3.21522$ & $4.92074$ & $4.01619$ & $3.31403$ \\
		\hline
		$\phi_{18}$ & $3.14159$ & $3.14159$ & $3.14159$ & $3.14159$ & $3.14159$ & $3.14159$ & $3.14159$ & $3.14159$ \\
		\hline
	\end{tabular}
	\caption{Parameters for the dynamically-corrected identity operation, the $x$ and $y$ rotations, and the $z$ rotation	by $\frac{\pi}{2}$ for a model with constant $J_i$.}
	\label{Tab:constantParamsI}
\end{table*}
\begin{table*}
	\centering
	\begin{tabular}{| c | c | c | c | c | c | c | c | c |}
		\hline
		Axis & $z$ & $z$ & $\hat{\vec{x}}+\hat{\vec{y}}$ & $\hat{\vec{x}}-\hat{\vec{y}}$ & $\hat{\vec{x}}+\hat{\vec{z}}$ & $\hat{\vec{x}}-\hat{\vec{z}}$ & $\hat{\vec{y}}+\hat{\vec{z}}$ & $\hat{\vec{y}}-\hat{\vec{z}}$ \\
		\hline
		Angle & $\pi$ & $3\pi/2$ & $\pi$ & $\pi$ & $\pi$ & $\pi$ & $\pi$ & $\pi$ \\
		\hline
		$j'^{\text{rot}}$ & $0.03333$ & $0.03333$ & $0.03333$ & $0.03333$ & $2.64575$ & $0.57485$ & $0.03333$ & $0.03333$ \\
		\hline
		$j_1^{\text{rot}}$ & $2.00000$ & $3.00000$ & $0.50509$ & $0.50509$ & $1.00000$ & $2.53276$ & $1.71541$ & $1.71541$ \\
		\hline
		$j_2^{\text{rot}}$ & $0.45510$ & $0.67658$ & $4.57172$ & $4.57172$ & --- & $0.03333$ & $0.13012$ & $0.13012$ \\
		\hline
		$j_3^{\text{rot}}$ & $2.00000$ & $3.00000$ & $0.50509$ & $0.50509$ & --- & $2.53276$ & $1.71541$ & $1.71541$ \\
		\hline
		$j_4^{\text{rot}}$ & --- & --- & --- & --- & --- & --- & --- & --- \\
		\hline
		$j_5^{\text{rot}}$ & --- & --- & --- & --- & --- & --- & --- & --- \\
		\hline
		$t_1^{\text{rot}}$ & $0.50027$ & $0.42475$ & $1.91411$ & $0.89008$ & $1.11072$ & $0.57686$ & $0.51303$ & $1.06916$ \\
		\hline
		$t_2^{\text{rot}}$ & $2.13931$ & $2.29036$ & $0.33565$ & $0.33565$ & --- & $1.56992$ & $1.55767$ & $1.55767$ \\
		\hline
		$t_3^{\text{rot}}$ & $0.50027$ & $0.42475$ & $0.89008$ & $1.91411$ & --- & $0.57686$ & $1.06916$ & $0.51303$ \\
		\hline
		$t_4^{\text{rot}}$ & --- & --- & --- & --- & --- & --- & --- & --- \\
		\hline
		$t_5^{\text{rot}}$ & --- & --- & --- & --- & --- & --- & --- & --- \\
		\hline
		$j_1$ & $14.8320$ & $6.08606$ & $12.9511$ & $2.98902$ & $9.36914$ & $5.01120$ & $6.21125$ & $5.20786$ \\
		\hline
		$j_2$ & $0.08568$ & $0.07273$ & $0.17844$ & $0.32786$ & $0.04832$ & $0.06079$ & $0.14018$ & $0.06111$ \\
		\hline
		$j_3$ & $17.4859$ & $5.41369$ & $1.75872$ & $10.2626$ & $8.17962$ & $8.18877$ & $6.64681$ & $5.73146$ \\
		\hline
		$j_4$ & $0.04172$ & $0.15263$ & $0.13529$ & $0.10200$ & $0.31474$ & $0.04798$ & $0.06600$ & $0.09549$ \\
		\hline
		$j_5$ & $10.6823$ & $11.8655$ & $2.87996$ & $12.5734$ & $8.40274$ & $4.43495$ & $6.24976$ & $4.24419$ \\
		\hline
		$j_6$ & $0.14942$ & $0.09036$ & $0.06331$ & $0.06508$ & $0.13463$ & $0.10140$ & $0.06618$ & $0.10577$ \\
		\hline
		$j_7$ & $18.4109$ & $7.12383$ & $11.4745$ & $11.8260$ & $4.71318$ & $8.08935$ & $8.75777$ & $5.42467$ \\
		\hline
		$j_8$ & $0.22390$ & $0.33033$ & $0.29352$ & $0.22181$ & $0.24977$ & $0.27460$ & $0.12328$ & $0.09174$ \\
		\hline
		$j_9$ & $19.9055$ & $6.73884$ & $7.19879$ & $10.0370$ & $9.68484$ & $1.04517$ & $7.22767$ & $3.56681$ \\
		\hline
		$j_{10}$ & $0.22414$ & $0.06294$ & $0.10262$ & $0.05338$ & $0.28410$ & $0.10052$ & $0.07137$ & $0.06428$ \\
		\hline
		$j_{11}$ & $13.7704$ & $10.7659$ & $3.49491$ & $14.3384$ & $6.16179$ & $7.94141$ & $6.30792$ & $8.25962$ \\
		\hline
		$j_{12}$ & $0.23365$ & $0.31156$ & $0.05490$ & $0.24371$ & $0.49621$ & $0.09732$ & $0.19631$ & $0.18793$ \\
		\hline
		$j_{13}$ & $21.2667$ & $8.21772$ & $12.9055$ & $2.82533$ & $7.59476$ & $5.90722$ & $8.13722$ & $2.81933$ \\
		\hline
		$j_{14}$ & $0.18113$ & $0.05662$ & $0.21813$ & $0.08250$ & $0.07428$ & $0.22768$ & $0.08147$ & $0.06690$ \\
		\hline
		$j_{15}$ & $10.3561$ & $5.26986$ & $3.93531$ & $12.2900$ & $14.9265$ & $2.69925$ & $4.29169$ & $4.53299$ \\
		\hline
		$j_{16}$ & $0.08696$ & $0.10441$ & $0.07098$ & $0.07438$ & $0.04470$ & $0.06062$ & $0.08320$ & $0.08701$ \\
		\hline
		$j_{17}$ & $6.22164$ & $5.54215$ & $16.6152$ & $9.95958$ & $13.3491$ & $5.46497$ & $3.83547$ & $6.02802$ \\
		\hline
		$j_{18}$ & $0.20041$ & $0.11415$ & $0.21540$ & $0.27869$ & $0.04341$ & $0.27100$ & $0.15614$ & $0.33781$ \\
		\hline
		$\phi_1$ & $1.75669$ & $0.49652$ & $3.09290$ & $3.85330$ & $3.10342$ & $4.02203$ & $5.43418$ & $0.98434$ \\
		\hline
		$\phi_2$ & $2.33422$ & $3.96681$ & $4.50049$ & $1.47780$ & $3.95197$ & $4.57343$ & $2.59242$ & $4.13376$ \\
		\hline
		$\phi_3$ & $5.30954$ & $2.49781$ & $2.81283$ & $4.59337$ & $0.93831$ & $2.09934$ & $3.69553$ & $2.48926$ \\
		\hline
		$\phi_4$ & $3.12939$ & $2.27810$ & $3.59551$ & $1.79992$ & $2.66582$ & $3.83942$ & $3.39603$ & $2.08419$ \\
		\hline
		$\phi_5$ & $3.28294$ & $4.08082$ & $5.27644$ & $3.92107$ & $4.21612$ & $3.34398$ & $1.55155$ & $4.53314$ \\
		\hline
		$\phi_6$ & $2.98278$ & $3.30503$ & $2.87973$ & $2.89507$ & $2.69878$ & $3.40785$ & $3.35411$ & $3.41282$ \\
		\hline
		$\phi_7$ & $2.97049$ & $2.86190$ & $3.01380$ & $3.17126$ & $4.52586$ & $2.99563$ & $0.90107$ & $1.87910$ \\
		\hline
		$\phi_8$ & $3.59125$ & $3.44407$ & $1.64479$ & $3.88246$ & $2.98595$ & $2.30693$ & $4.03589$ & $4.29502$ \\
		\hline
		$\phi_9$ & $1.48541$ & $3.47186$ & $5.16677$ & $2.82468$ & $3.74931$ & $2.66493$ & $1.96720$ & $2.63959$ \\
		\hline
		$\phi_{10}$ & $3.27256$ & $2.84195$ & $3.83777$ & $3.36406$ & $3.60174$ & $5.02774$ & $2.53451$ & $3.23638$ \\
		\hline
		$\phi_{11}$ & $5.26209$ & $5.76930$ & $2.23586$ & $4.49378$ & $0.60704$ & $2.67576$ & $5.92522$ & $4.83535$ \\
		\hline
		$\phi_{12}$ & $3.40959$ & $2.13972$ & $3.66723$ & $2.42232$ & $3.33144$ & $2.96314$ & $2.81749$ & $1.66024$ \\
		\hline
		$\phi_{13}$ & $3.45772$ & $4.43079$ & $0.72011$ & $1.64091$ & $2.42357$ & $2.97994$ & $4.50915$ & $4.35308$ \\
		\hline
		$\phi_{14}$ & $2.92306$ & $2.96829$ & $4.41970$ & $3.49835$ & $3.12093$ & $2.89307$ & $3.28735$ & $3.44457$ \\
		\hline
		$\phi_{15}$ & $3.57310$ & $3.28504$ & $3.75495$ & $2.01767$ & $4.37853$ & $3.39137$ & $2.10098$ & $2.08324$ \\
		\hline
		$\phi_{16}$ & $3.08844$ & $2.50621$ & $3.16872$ & $3.79660$ & $3.71673$ & $3.13385$ & $3.66449$ & $2.83612$ \\
		\hline
		$\phi_{17}$ & $2.42464$ & $3.66783$ & $4.04567$ & $2.14381$ & $1.88580$ & $0.77083$ & $2.15351$ & $3.94458$ \\
		\hline
		$\phi_{18}$ & $3.14159$ & $3.14159$ & $3.14159$ & $3.14159$ & $3.14159$ & $3.14159$ & $3.14159$ & $3.14159$ \\
		\hline
	\end{tabular}
	\caption{Parameters for the dynamically-corrected $z$ rotations	by $\pi$ and $3\pi/2$, and for the rotations by
		$\hat{\vec{x}}\pm\hat{\vec{y}}$, $\hat{\vec{x}}\pm\hat{\vec{z}}$, and $\hat{\vec{y}}\pm\hat{\vec{z}}$ for a model with constant $J_i$.}
	\label{Tab:constantParamsII}
\end{table*}
\begin{table*}
	\centering
	\begin{tabular}{| c | c | c | c | c | c | c | c | c |}
		\hline
		Axis & $\hat{\vec{x}}+\hat{\vec{y}}+\hat{\vec{z}}$ & $\hat{\vec{x}}+\hat{\vec{y}}+\hat{\vec{z}}$ & $-\hat{\vec{x}}+\hat{\vec{y}}+\hat{\vec{z}}$ & $-\hat{\vec{x}}+\hat{\vec{y}}+\hat{\vec{z}}$ & $\hat{\vec{x}}-\hat{\vec{y}}+\hat{\vec{z}}$ & $\hat{\vec{x}}-\hat{\vec{y}}+\hat{\vec{z}}$ & $\hat{\vec{x}}+\hat{\vec{y}}-\hat{\vec{z}}$ & $\hat{\vec{x}}+\hat{\vec{y}}-\hat{\vec{z}}$ \\
		\hline
		Angle & $2\pi/3$ & $4\pi/3$ & $2\pi/3$ & $4\pi/3$ & $2\pi/3$ & $4\pi/3$ & $2\pi/3$ & $4\pi/3$ \\
		\hline
		$j'^{\text{rot}}$ & $0.03333$ & $0.03333$ & $0.03333$ & $0.03333$ & $0.03333$ & $0.03333$ & $0.03333$ & $0.03333$ \\
		\hline
		$j_1^{\text{rot}}$ & $0.56170$ & $0.74456$ & $0.32145$ & $0.29210$ & $0.56170$ & $0.74456$ & $0.29210$ & $0.32145$ \\
		\hline
		$j_2^{\text{rot}}$ & $2.41400$ & $3.22657$ & $14.0228$ & $8.36716$ & $2.41400$ & $3.22657$ & $8.36716$ & $14.0228$ \\
		\hline
		$j_3^{\text{rot}}$ & $0.56170$ & $0.74456$ & $0.32145$ & $0.29210$ & $0.56170$ & $0.74456$ & $0.29210$ & $0.32145$ \\
		\hline
		$j_4^{\text{rot}}$ & --- & --- & --- & --- & --- & --- & --- & --- \\
		\hline
		$j_5^{\text{rot}}$ & --- & --- & --- & --- & --- & --- & --- & --- \\
		\hline
		$t_1^{\text{rot}}$ & $2.21343$ & $1.97092$ & $1.17574$ & $1.18212$ & $0.52564$ & $0.54891$ & $1.83346$ & $1.81513$ \\
		\hline
		$t_2^{\text{rot}}$ & $0.40078$ & $0.62001$ & $0.14898$ & $0.12427$ & $0.40078$ & $0.62001$ & $0.12427$ & $0.14898$ \\
		\hline
		$t_3^{\text{rot}}$ & $0.52564$ & $0.54891$ & $1.81513$ & $1.83346$ & $2.21343$ & $1.97092$ & $1.18212$ & $1.17574$ \\
		\hline
		$t_4^{\text{rot}}$ & --- & --- & --- & --- & --- & --- & --- & --- \\
		\hline
		$t_5^{\text{rot}}$ & --- & --- & --- & --- & --- & --- & --- & --- \\
		\hline
		$j_1$ & $10.7661$ & $11.5740$ & $5.00930$ & $5.44077$ & $10.2269$ & $18.7776$ & $5.90484$ & $7.15782$ \\
		\hline
		$j_2$ & $0.11510$ & $0.28828$ & $0.09538$ & $0.29018$ & $0.11723$ & $0.05318$ & $0.31409$ & $0.17930$ \\
		\hline
		$j_3$ & $8.89117$ & $3.65918$ & $19.1213$ & $11.3119$ & $10.2021$ & $10.0969$ & $4.31741$ & $7.46876$ \\
		\hline
		$j_4$ & $0.20559$ & $0.05901$ & $0.09606$ & $0.04124$ & $0.09354$ & $0.15789$ & $0.06110$ & $0.05518$ \\
		\hline
		$j_5$ & $4.83475$ & $25.6828$ & $16.7694$ & $9.32045$ & $8.16447$ & $22.2191$ & $6.51526$ & $5.27719$ \\
		\hline
		$j_6$ & $0.13395$ & $0.06718$ & $0.05482$ & $0.43300$ & $0.11700$ & $0.17654$ & $0.08278$ & $0.05364$ \\
		\hline
		$j_7$ & $4.48507$ & $3.05551$ & $5.41672$ & $17.3620$ & $11.0271$ & $3.90888$ & $6.85651$ & $17.3085$ \\
		\hline
		$j_8$ & $0.04527$ & $0.04005$ & $0.26657$ & $0.17560$ & $0.21473$ & $0.32907$ & $0.18002$ & $0.07704$ \\
		\hline
		$j_9$ & $2.18431$ & $6.29185$ & $4.22555$ & $3.67047$ & $12.0647$ & $21.4374$ & $0.55896$ & $13.0452$ \\
		\hline
		$j_{10}$ & $0.04490$ & $0.20990$ & $0.50948$ & $0.50657$ & $0.07343$ & $0.39065$ & $0.86102$ & $0.13407$ \\
		\hline
		$j_{11}$ & $17.2965$ & $5.91075$ & $18.5654$ & $18.3140$ & $8.36832$ & $20.4086$ & $7.40366$ & $15.0703$ \\
		\hline
		$j_{12}$ & $0.04339$ & $0.06178$ & $0.03382$ & $0.11821$ & $0.27946$ & $0.04014$ & $0.13026$ & $0.05362$ \\
		\hline
		$j_{13}$ & $6.82614$ & $27.2096$ & $19.6292$ & $6.85666$ & $6.94247$ & $26.3145$ & $7.57882$ & $6.28250$ \\
		\hline
		$j_{14}$ & $0.14376$ & $0.06382$ & $0.34693$ & $0.30954$ & $0.16238$ & $0.16546$ & $0.10877$ & $0.19997$ \\
		\hline
		$j_{15}$ & $5.06561$ & $4.64735$ & $4.87618$ & $15.9969$ & $7.65720$ & $9.27366$ & $4.90249$ & $5.97533$ \\
		\hline
		$j_{16}$ & $0.22779$ & $0.05061$ & $0.03903$ & $0.04383$ & $0.07173$ & $0.13057$ & $0.06292$ & $0.05197$ \\
		\hline
		$j_{17}$ & $7.64291$ & $5.61436$ & $26.8249$ & $12.2456$ & $6.38671$ & $20.9800$ & $6.40504$ & $6.57259$ \\
		\hline
		$j_{18}$ & $0.29524$ & $0.36538$ & $0.51048$ & $0.70207$ & $0.22653$ & $0.26497$ & $0.46266$ & $0.07298$ \\
		\hline
		$\phi_1$ & $1.86227$ & $2.26580$ & $1.99571$ & $2.43151$ & $3.12683$ & $3.13524$ & $2.02942$ & $4.30594$ \\
		\hline
		$\phi_2$ & $2.39636$ & $1.49241$ & $3.47874$ & $2.99874$ & $3.84944$ & $3.11065$ & $3.65426$ & $3.40109$ \\
		\hline
		$\phi_3$ & $3.09316$ & $3.37813$ & $2.91187$ & $3.89836$ & $1.93029$ & $3.27606$ & $3.38573$ & $2.83522$ \\
		\hline
		$\phi_4$ & $4.08038$ & $3.08759$ & $2.24376$ & $3.35420$ & $2.90934$ & $3.18770$ & $3.04261$ & $3.13690$ \\
		\hline
		$\phi_5$ & $1.96027$ & $3.05081$ & $5.45630$ & $4.98881$ & $4.25714$ & $1.86445$ & $3.74543$ & $2.14373$ \\
		\hline
		$\phi_6$ & $3.21424$ & $3.14847$ & $4.07229$ & $2.53448$ & $3.14612$ & $2.39157$ & $2.14569$ & $2.56715$ \\
		\hline
		$\phi_7$ & $4.16769$ & $3.99172$ & $2.41585$ & $2.77336$ & $4.61272$ & $4.62757$ & $4.16869$ & $4.07547$ \\
		\hline
		$\phi_8$ & $2.17830$ & $3.35751$ & $2.70266$ & $3.66705$ & $2.39626$ & $3.51918$ & $2.41949$ & $3.83791$ \\
		\hline
		$\phi_9$ & $3.57785$ & $5.37022$ & $3.51253$ & $1.23518$ & $4.24717$ & $1.58182$ & $4.95506$ & $2.07100$ \\
		\hline
		$\phi_{10}$ & $2.70602$ & $1.22627$ & $1.69335$ & $3.06442$ & $3.74150$ & $4.54924$ & $1.50139$ & $2.59654$ \\
		\hline
		$\phi_{11}$ & $1.35997$ & $5.20760$ & $3.18859$ & $2.74756$ & $0.31262$ & $2.40164$ & $2.38693$ & $5.09851$ \\
		\hline
		$\phi_{12}$ & $4.48065$ & $2.96135$ & $2.94189$ & $3.71204$ & $3.34590$ & $3.69582$ & $4.17284$ & $3.64342$ \\
		\hline
		$\phi_{13}$ & $2.31352$ & $2.31340$ & $4.59434$ & $2.49796$ & $1.86224$ & $1.83430$ & $2.89132$ & $4.21655$ \\
		\hline
		$\phi_{14}$ & $3.43885$ & $2.63958$ & $3.43121$ & $3.79832$ & $2.96260$ & $3.20157$ & $3.45387$ & $2.74168$ \\
		\hline
		$\phi_{15}$ & $3.35857$ & $2.41218$ & $2.67771$ & $2.05334$ & $3.99016$ & $2.92032$ & $2.11405$ & $2.92801$ \\
		\hline
		$\phi_{16}$ & $3.31546$ & $3.96840$ & $2.93719$ & $2.76419$ & $2.94494$ & $2.73833$ & $3.89997$ & $3.22086$ \\
		\hline
		$\phi_{17}$ & $2.51257$ & $4.36821$ & $2.44037$ & $6.09444$ & $3.72074$ & $6.00654$ & $1.43218$ & $3.49669$ \\
		\hline
		$\phi_{18}$ & $3.14159$ & $3.14159$ & $3.14159$ & $3.14159$ & $3.14159$ & $3.14159$ & $3.14159$ & $3.14159$ \\
		\hline
	\end{tabular}
	\caption{Parameters for the dynamically-corrected $\hat{\vec{x}}\pm\hat{\vec{y}}\pm\hat{\vec{z}}$ rotations  for a model with constant $J_i$.}
	\label{Tab:constantParamsIII}
\end{table*}

\end{document}